\documentclass[]{pasj01}
\usepackage{natbib,bm,url}
\begin{document} 
\Received{}
\Accepted{}

\title{SIRIUS project. II. a new tree-direct hybrid code for smoothed particle hydrodynamics/$N$-body simulations of star clusters}

\author{Michiko S. \textsc{Fujii}\altaffilmark{1}%
}
\altaffiltext{1}{Department of Astronomy, Graduate School of Science, The University of Tokyo, 7-3-1 Hongo, Bunkyo-ku, Tokyo 113-0033, Japan}
\email{fujii@astron.s.u-tokyo.ac.jp}

\author{Takayuki R. \textsc{Saitoh}\altaffilmark{2, 3}}
\altaffiltext{2}{Department of Planetology, Graduate School of Science, Kobe University, 1-1 Rokkodai-cho, Nada-ku, Kobe, Hyogo 657-8501, Japan}
\altaffiltext{3}{Earth-Life Science Institute, Tokyo Institute of Technology, 2-12-1 Ookayama, Meguro-ku, Tokyo 152-8551, Japan}
\email{saitoh@people.kobe-u.ac.jp}

\author{Long \textsc{Wang}\altaffilmark{1}%
\thanks{JSPS Fellow}
}
\email{long.wang@astron.s.u-tokyo.ac.jp}

\author{Yutaka \textsc{Hirai}\altaffilmark{4, 5*}}
\altaffiltext{4}{Astronomical Institute, Tohoku University, 6-3, Aramaki, Aoba-ku, Sendai, Miyagi 980-8578, Japan}
\altaffiltext{5}{RIKEN Center for Computational Science, 7-1-26 Minatojima-minami-machi, Chuo-ku,
Kobe, Hyogo 650-0047, Japan}
\email{yutaka.hirai@astr.tohoku.ac.jp}


\KeyWords{methods: numerical --- stars: formation --- ISM: clouds --- open clusters and associations: general --- galaxies: star clusters: general} 

\maketitle

\begin{abstract}
Star clusters form via clustering star formation inside molecular clouds. In order to understand the dynamical evolution of star clusters in their early phase, in which star clusters are still embedded in their surrounding gas, we need an accurate integration of individual stellar orbits without gravitational softening in the systems including both gas and stars, as well as modeling individual stars with a realistic mass function. 
We develop a new tree-direct hybrid smoothed particle hydrodynamics/$N$-body code, \textsc{ASURA+BRIDGE}, in which stars are integrated using a direct $N$-body scheme or \textsc{PeTar}, a particle-particle particle-tree scheme code, without gravitational softening. In \textsc{ASURA+BRIDGE}, stars are assumed to have masses randomly drawn from a given initial mass function. With this code, we perform star-cluster formation simulations starting from molecular clouds without gravitational softening. We find that artificial dense cores in star-cluster centers due to the softening disappear when we do not use softening. We further demonstrate that star clusters are built up via mergers of smaller clumps. Star clusters formed in our simulations include some dynamically formed binaries with the minimum semi-major axes of a few au, and the binary fraction is higher for more massive stars.
\end{abstract}

\section{Introduction}

Star clusters form via collective formation of stars in giant molecular clouds (GMCs). Recent numerical studies have shown that star formation in turbulent molecular clouds successfully results in the formation of star clusters \citep[][and references therein]{2003MNRAS.343..413B,2009MNRAS.392..590B,2012A&A...542A..49B,2018ApJ...859...68K,2018NatAs...2..725H,2019ARA&A..57..227K,2019MNRAS.489.1880H,2020MNRAS.497.3830F}.

Most of the simulations for star cluster formation are performed using sink particles for the star formation \citep{1995MNRAS.277..362B,2003MNRAS.343..413B,2005MNRAS.356.1201B}. With this method, sink particles are formed as seeds of stars when the local density exceeds a threshold density. The sink particles evolve by eating gas accreting into a sink radius. In order to resolve low-mass stars, the resolution of the simulations must be sufficiently high. In a smoothed-particle hydrodynamics (SPH) simulation using a sink method \citep{2012MNRAS.419.3115B}, in which the formation of brown dwarfs is resolved, the mass resolution of gas particles was $\sim 10^{-5}M_{\odot}$. In this simulation, the total gas mass was only $500M_{\odot}$. 
The limit of total gas mass necessary for resolving individual stars is similar for the adaptive mesh refinement (AMR) hydrodynamics code. \citet{2011ApJ...740...74K} performed a simulation of star cluster formation using an AMR code with sink particles resolving down to $0.1M_{\odot}$, and the total gas mass was $1000\,M_{\odot}$. Thus, the mass of forming star clusters is limited to $\sim 1000 M_{\odot}$ if we use sink methods to resolve individual stars down to $0.1M_{\odot}$.

Although the typical mass of open clusters in the Milky Way galaxy is $\sim 1000 M_{\odot}$ and the mass of embedded clusters are even smaller \citep{2003ARA&A..41...57L}, young massive clusters such as NGC 3603 have masses of $>10^4 M_{\odot}$ \citep[][and the references therein]{2010ARA&A..48..431P}. In order to simulate the formation of such massive star clusters, recent numerical studies have applied sink methods for `cluster particles,' in which individual stellar particles assumed to contain some stars with a given mass spectrum in star-cluster-scale simulations \citep{2005MNRAS.359..809C,2018ApJ...859...68K,2018NatAs...2..725H,2019MNRAS.489.1880H,2020MNRAS.497.3830F} and in galaxy simulations \citep{2014MNRAS.445..581H,2017MNRAS.466.1903G,2018ApJ...853..173K}. Such a super-particle approach can treat massive clusters, but the dynamical evolution of star clusters which is driven by collisional encounters of stars cannot be included.

Instead of using sink methods, we probabilistically form stars following a given mass function. In \citet[][hereafter Paper I]{2020arXiv200512906H}, we demonstrated that our star formation scheme can reproduce observed stellar mass function. There are some previous works that adopted similar methods. In \citet{2019ApJ...887...62W}, stellar masses initially drawn from a given mass function are assigned to sink particles and form stars every time when the sink mass exceeds the assigned stellar mass. \citet{2013MNRAS.435.1701C} assumed a probabilistic star formation similar to ours.
For a simulation of dwarf galaxies, \citet{2020ApJ...891....2L} also adopted a probabilistic star-formation scheme. With these star formation schemes, stellar distribution with a realistic stellar mass function can be realized without resolving gas particles down to $10^{-5} M_{\odot}$.

If we can resolve individual stars, we are able to follow the dynamical evolution of star clusters with a mass spectrum. With the mass spectrum, massive stars immediately sink to the cluster center and form hard binaries. Such binaries kick out their surrounding stars and form runaway stars \citep{2011Sci...334.1380F,2012ApJ...746...15B}. Recent simulations suggest that resolving individual stars make the photoionization time-scale longer compared with a cluster particle \citep{2020MNRAS.tmp.2473D}. In addition, runaway stars may play an important role in the heating out of molecular clouds \citep{2020MNRAS.494.3328A}. If massive stars are ejected from the center or merge via few-body interactions, the heating process can dramatically change. This may trigger the formation of multiple stellar populations in the Orion Nebula Cluster \citep{Kroupa2018,Wang2019} and in globular clusters \citep{Wang2020a}. Thus, integration of individual orbits of stars using an accurate integrator without gravitational softening is important to understand the formation process of star clusters.

There are some simulations of star clusters coupled with gas dynamics without gravitational softening for stars. 
\citet{2019ApJ...887...62W}  coupled a fourth-order Hermite integrator, \textsc{ph4}, \citep{1992PASJ...44..141M} with an AMR code, \textsc{FLASH}, using the Astrophysical Multi-purpose Software Environment \citep[\textsc{AMUSE},][]{2013CoPhC.183..456P, 2013A&A...557A..84P,AMUSE}. AMUSE is a Python framework and provides us an interface to couple existing codes (community codes) written in different programming languages. They implemented the Bridge scheme \citep{2007PASJ...59.1095F}, which is an enhanced mixed variable symplectic scheme, on AMUSE \citep{2012MNRAS.420.1503P}. In this scheme,  gravitational force from gas to stars is given in a fixed timestep as a velocity kick. They performed simulations of star-cluster formation up to $\sim 1000$ stars \citep{2020arXiv200309011W}.  
Using AMUSE, \citet{2012MNRAS.420.1503P} also performed simulations of star clusters embedded in a molecular cloud. Although they did not assume star formation during the simulations, they included stellar feedback due to stellar wind and supernova. In this simulation, 1000 stars are also used. 

\citet{2020MNRAS.tmp.2473D} implemented a direct $N$-body code combined with \textsc{FLASH} without using AMUSE. They used \textsc{NBODY6}, which is also a fourth-order Hermite code including regularization scheme for binaries \citep{2003gnbs.book.....A}. By performing embedded cluster simulations with feedback from massive stars using maximum $\sim 400$ stars, they showed that the gas-expulsion time-scale becomes longer in star-by-star simulations compared with that in simulations with cluster-particles \citep{2020MNRAS.tmp.2473D}. Thus, star-by-star treatment without gravitational softening is important for performing simulations of star cluster formation and evolution.

In these previous studies, however, the application was limited for typical open clusters. 
For larger simulations such as young massive clusters and globular clusters, we develop a new code combining highly accurate $N$-body integrators with an $N$-body/SPH code, \textsc{ASURA}, using the Bridge scheme \citep{2007PASJ...59.1095F}. 
The $N$-body integrators can be chosen from the sixth-order Hermite scheme \citep{2008A&A...483..171N} or the \textsc{PeTar} code, which uses the Particle-Particle Particle-Tree (P$^3$T) and the slow-down algorithmic regularization (SDAR) methods, \citep{2011PASJ...63..881O,2015ComAC...2....6I,Wang2020b,Wang2020c}.
With our new code, \textsc{ASURA+BRIDGE}, we first aim to perform simulations of massive star clusters with $>10^4 M_{\odot}$. We will further extend this scheme up to globular-cluster and even to galaxy-scale simulations including chemo-dynamical evolution \citep[e.g.,][]{2017AJ....153...85S, 2019ApJ...885...33H} as the SIRIUS, SImulations Resolving IndividUal Stars, project.

\section{Numerical method}
In this section, we describe our new code, \textsc{ASURA+BRIDGE}. We implement the Bridge scheme \citep{2007PASJ...59.1095F} in an $N$-body/SPH simulation code, \textsc{ASURA} \citep{2008PASJ...60..667S,2009PASJ...61..481S}. 

\subsection{\textsc{ASURA}: $N$-body/smoothed-particle hydrodynamics code}
\textsc{ASURA} is an $N$-body/SPH code that was originally developed for simulations of galaxy formation \citep{2008PASJ...60..667S, 2009PASJ...61..481S}. 
In \textsc{ASURA}, the gravitational interactions are solved by the tree 
with GRAPE methods. The current version of \textsc{ASURA} uses Phantom-GRAPE \citep{2012NewA...17...82T, 2013NewA...19...74T}, a software emulating GRAPE \citep{1990Natur.345...33S,2003PASJ...55.1163M,2005PASJ...57.1009F}, and thus can work on ordinary supercomputers. Unlike star cluster simulations, the required force accuracy in galaxy formation is not so high because of their collisionless nature. We introduce gravitational softening by changing the force law as the Plummer model. To treat a system consisting of different gravitational softening, \textsc {ASURA} adopts the symmetrized Plummer potential model for force evaluation \citep{2012NewA...17...76S}.
The hydrodynamical interactions are solved with the SPH method \citep{1977AJ.....82.1013L,1977MNRAS.181..375G}. 
The original version of SPH adopts density as a fundamental quantity and therefore it is difficult to handle the fluid instabilities induced at contact discontinuities. 
\textsc{ASURA}, in contrast, adopts the density-independent SPH  \citep[DISPH, ][]{2013ApJ...768...44S}, which uses pressure as a fundamental quantity and can solve the growth of fluid instabilities.

The time-integration is carried out with the second-order symplectic integrator, the leap-frog method. In order to accelerate the integration, \textsc{ASURA} equips an individual and block timestep method \citep{1986LNP...267..156M, 1991PASJ...43..859M} and uses the FAST method \citep{2010PASJ...62..301S}. 
Integration errors induced by the individual timestep method in the SPH part are reduced by limiting the step-size difference among the nearest particles within a factor of four \citep{2009ApJ...697L..99S}.

\textsc{ASURA} also deals with radiative cooling and heating of the interstellar medium with Cloudy ver.13.05 \citep{1998PASP..110..761F, 2013RMxAA..49..137F, 2017RMxAA..53..385F}, star formation from dense and cold gas, and energy and metal release from stars. For the star formation scheme, we describe more details in Section 2.4.

\textsc{ASURA} is parallelized using the message passing library (MPI). The parallelization strategy of \textsc{ASURA} is the same as that of  \citet{2004PASJ...56..521M}. Thus, \textsc{ASURA} can work with arbitrary numbers of MPI processes.

\subsection{\textsc{ASURA+BRIDGE}: a hybrid integrator}
We implement the Bridge scheme into \textsc{ASURA}. The Bridge scheme is an extension of a mixed-variable symplectic integrator \citep{1991AJ....102.1528W,1991CeMDA..50...59K}. In \citet{1991AJ....102.1528W}, the Hamiltonian of the system (planets orbiting around the sun) was split into a Kepler motion around the sun and perturbations from other planets. In every timestep, the motions of planets are integrated as Kepler motions perturbed by the other planets, and the perturbation was calculated using an $N$-body scheme. 

In \citet{2007PASJ...59.1095F}, the Hamiltonian splitting was applied for a star cluster embedded in its host galaxy, in which both the star cluster and galaxy were modeled as $N$-body systems. In this original Bridge scheme, the motions of stars in the star cluster were integrated using a fourth-order Hermite scheme \citep{1992PASJ...44..141M}, and every fixed shared timestep (hereafter Bridge timestep), the stars in the star cluster are perturbed by the gravitational force from the galaxy particles. The force among galaxy particles and galaxy-star cluster particles are calculated using a tree code \citep{1986Natur.324..446B}. While the galaxy is integrated using a second-order leap-frog integrator with a shared timestep (Bridge timestep), star-cluster particles are integrated using a fourth-order Hermite scheme with individual block timesteps. This allows us to calculate galaxy particles fast enough and star-cluster particles accurate enough without gravitational softening. With the Bridge scheme, we can follow the dynamical evolution of star clusters as a collisional system inside a live galaxy \citep{2008ApJ...686.1082F}. The integrator for star clusters was later upgraded to a sixth-order Hermite scheme \citep{2008NewA...13..498N, 2009ApJ...695.1421F,2010ApJ...716L..80F}.

\citet{2012MNRAS.420.1503P} further enhanced the Bridge scheme to an $N$-body/SPH simulation in the framework of \textsc{AMUSE}. They integrated SPH particles using an SPH code, \textsc{Fi}, and stellar particles using a fourth-order Hermite code, \textsc{ph4}. Every Bridge timestep ($\Delta t_{\rm B}$), stellar particles are perturbed by velocity kicks from gas particles and integrated during a Bridge step using \textsc{ph4}. Gas particles are on the other hand integrated using \textsc{Fi}. The data are transferred to each other in the framework of \textsc{AMUSE}.  \textsc{AMUSE} automatically changes the scaling for one code to another, but the data copy becomes a bottleneck when we perform a large simulation. We therefore develop a new code, \textsc{ASURA+BRIDGE}, without using \textsc{AMUSE} framework. 

In every Bridge timestep, \textsc{ASURA} sends all stellar data to the Bridge part (direct integration part). The Bridge part integrates all stellar particles during $\Delta t_{\rm B}$ using a sixth-order Hermite scheme \citep{2008NewA...13..498N} with individual block timesteps. During $\Delta t_{\rm B}$ gas particles are also integrated using block timesteps independent of stellar particles using \textsc{ASURA}, but the most of gas particle has a stepsize larger than the Bridge step. \textsc{ASURA+BRIDGE} can set a softening length for stellar particles different from that of gas particles, and integration without softening is also possible. This enables us to treat star clusters as collisional systems.

\subsection{\textsc{PeTar}}
For the direct integration, we also equip \textsc{PeTar} \citep{Wang2020c}.
\textsc{PeTar} is a tree-direct hybrid $N$-body code that combines the P$^3$T scheme  \citep{2011PASJ...63..881O, 2015ComAC...2....6I} with the SDAR scheme \citep{Wang2020b}. 
The P$^3$T scheme separates the interactions in an $N$-body system into two parts, long- and short-ranges, via a Hamiltonian splitting.
For each particle, the long-range force contributes less but the calculation cost is the dominant part in computing because we need to calculate the force from all the other particles. 
Thus, an efficient and less-accurate method, such as the particle-tree method with the second-order leap-frog integrator, can be used for the long-range force.
On the other hand, the short-range interactions from only a few neighbor particles can be handled by an expensive but accurate direct $N$-body method, such as the forth-order Hermite integrator with individual (block) timesteps.
Compared to pure direct $N$-body methods, the P$^3$T scheme can significantly reduce the computational cost from $O(N^2)$ to $O(N\log{N})$ while keeping a sufficient accuracy for close encounters of stars.

In star clusters, in addition, multiple systems such as binaries and triples play very important roles in the dynamical evolution of the system. Integrating them properly is important to follow the long-term evolution of star clusters and the formation and evolution of binaries in star clusters.
The Hermite integrator is not always sufficient to handle such multiple systems. 
When binaries are in a highly eccentric orbit, the numerical error can significantly increase during the peri-center passage.  
After a long-term evolution, the accumulated error becomes non-negligible and results in the wrong orbits of binaries.
The direct integration of multiple systems is also very time-consuming since their timesteps required in order to maintain enough accuracy are much smaller than those for the entire system. 
The SDAR scheme \citep{Wang2020b}, which is a time-transformed explicit symplectic integrator with the slowdown method, solves this issue, and it is implemented in \textsc{PeTar}.

In order to have a high-performance with parallel computing, \textsc{PeTar} uses the parallelization framework for developing particle simulation codes (\textsc{fdps}) \citep{Iwasawa2016,Namekata2018}.
The GPU and SIMD (AVX, AVX2 and AVX512) acceleration for force calculations are also implemented.
Thus, \textsc{PeTar} is a high-performance $N$-body code that can handle the large-scale simulations of massive star clusters including binaries and multiples.

\subsection{\textsc{CELib}: star formation scheme}
We use a library \textsc{CELib} \citep{2017AJ....153...85S} for the star formation. The details of this star-formation scheme are described in \citet{2020arXiv200512906H}. Here we briefly describe the scheme of star formation, which is a probabilistic star formation often adopted for galaxy simulations. 

For the formation of star particles from gas particles, we require three conditions of gas particles: (1) the density is higher than the threshold density ($n_{\rm{th}}$), (2) the temperature is lower than the threshold temperature ($T_{\rm{th}}$), and (3) gas particles are converging ($\nabla\cdot\bm{v}<0$). 
Gas particles satisfying these conditions are probabilistically selected to form new star particles. The probability is determined using local gas density and the star formation efficiency per free-fall time. The star formation efficiency is a parameter, and we set it as 0.02 for our fiducial value. This value is consistent with the observed star formation efficiency in a molecular cloud \citep[e.g.,][]{2019ARA&A..57..227K}.

Once a gas particle becomes eligible to form a star particle, we assign a stellar mass to the particle. This stellar mass is 
sampled following a given initial mass function (IMF) using the chemical evolution library \citep[\textsc{CELib},][]{2017AJ....153...85S}. In this study, we assume the Kroupa IMF \citep{2001MNRAS.322..231K} with a lower- and upper-mass limit of 0.1 and 150\,$M_{\odot}$. After the assignment of mass, a stellar particle is generated by assembling mass from the surrounding gas particles. For the mass assemblage, we first determine the region that contains the mass of 5 to 10 times the mass of the stellar particle. We then set the maximum search radius ($r_{\rm{max}}$) of 0.2 pc to assemble masses of gas particles. If the mass in this region does not reach half of the forming stellar mass, we re-assign the stellar mass to the new star particle. \citet{2020arXiv200512906H} confirmed that $r_{\rm{max}}$ = 0.2 pc is enough to fully sample Kroupa IMF from 0.1 to 150 $M_{\odot}$. Even if we set the maximum mass of 150\,$M_{\odot}$, such massive stars may not be able to form due to the lack of the surrounding gas. This naturally reproduces the maximum mass limit depending on the total stellar mass of star clusters seen in observations \citep{2006MNRAS.365.1333W}.
The forming stars take over the positions and velocities from the parental gas particles.

\section{Results: Test simulations}
Here we show the results of our test calculation using the new code. We perform two kinds of star cluster simulations: the embedded cluster model and the star cluster formation model. The former is similar to a model used in \citet{2012MNRAS.420.1503P}, in which stars initially embedded in a molecular cloud. 
As a star cluster formation model, we simulate a turbulent molecular cloud \citep{2003MNRAS.343..413B} with a star formation scheme using \textsc{CELib} star-by-star mode as described in Paper I. 

\subsection{Test 1: Embedded cluster model}

We first test an embedded cluster model, a star cluster initially embedded in a gas sphere. This model is similar to one of the models in \citet{2012MNRAS.420.1503P}. We perform simulations using both \textsc{ASURA} and \textsc{ASURA+BRIDGE} and compare the results.

\subsubsection{Initial conditions}
We adopt a model similar to model A2 in \citet{2012MNRAS.420.1503P}. In this model, both gas and stellar particles distribute following a Plummer model \citep{1911MNRAS..71..460P} with the scale length of $a=0.5$\,pc. The stellar mass fraction to the total mass (i.e., star formation efficiency) is set to be 0.3.

\begin{figure*}
 \begin{center}
  \includegraphics[width=5.2cm]{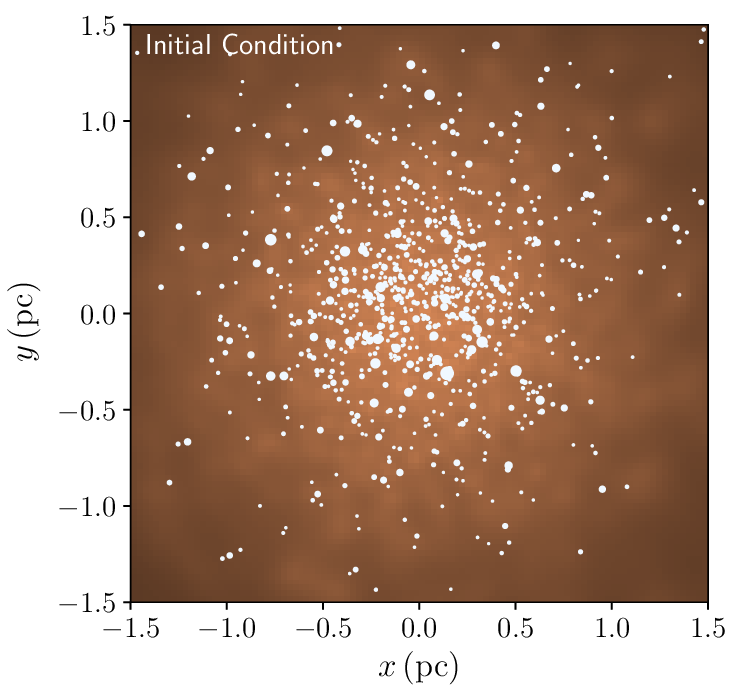}
  \includegraphics[width=5.2cm]{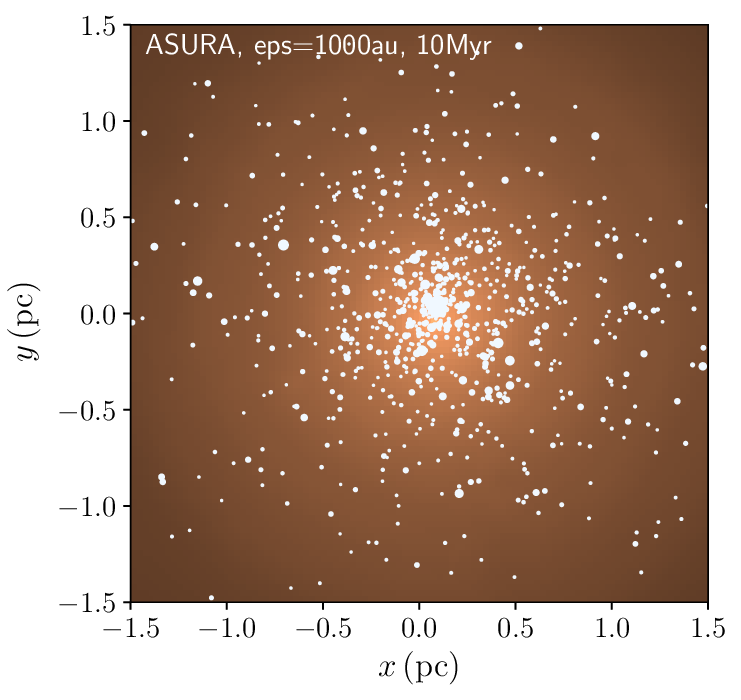}
  \includegraphics[width=5.2cm]{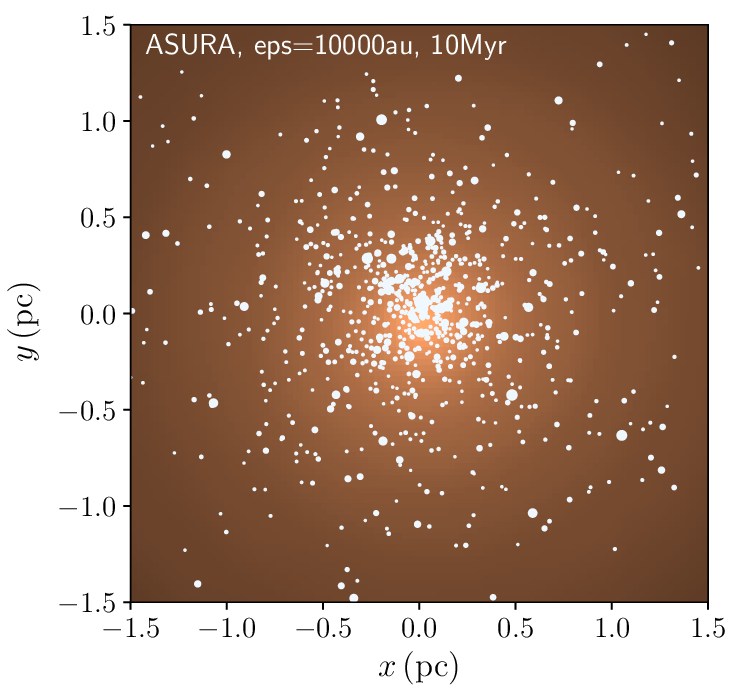}\\
  \includegraphics[width=5.2cm]{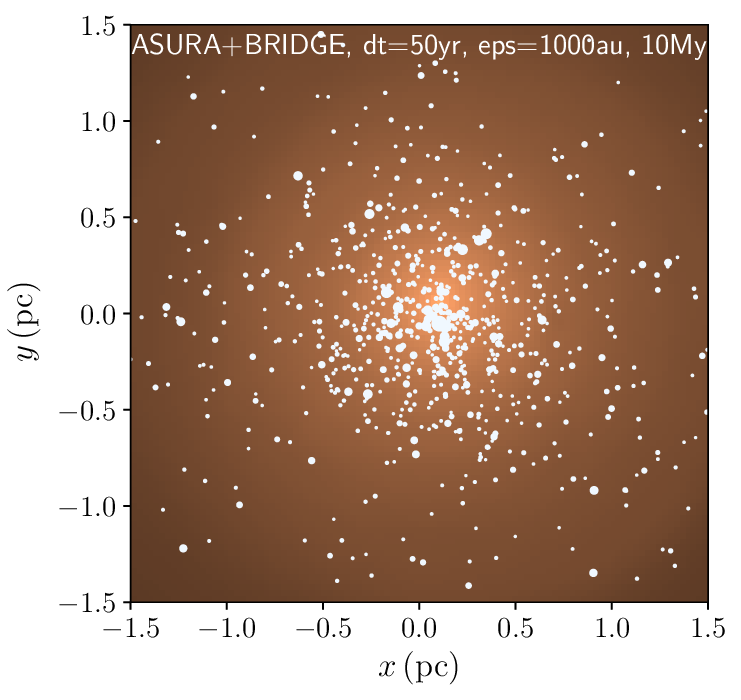}
  \includegraphics[width=5.2cm]{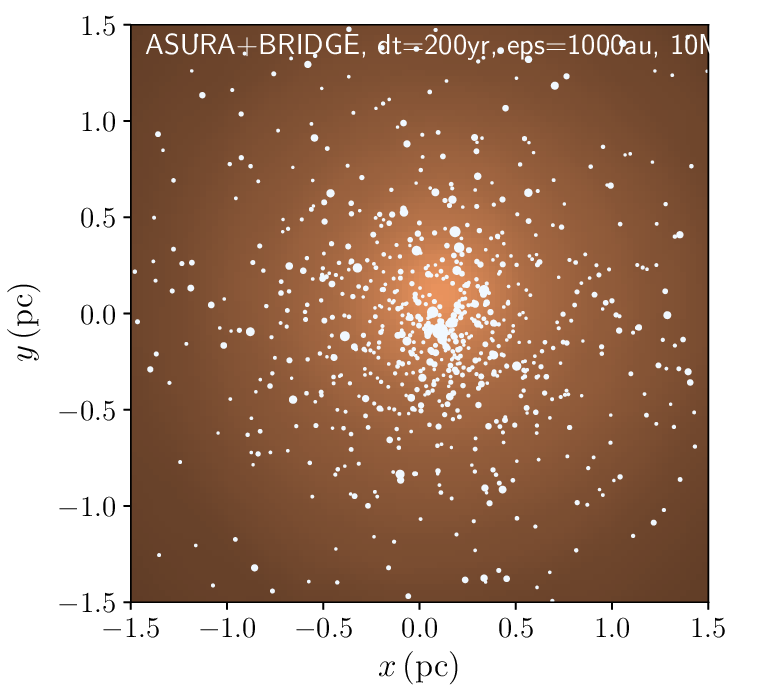}
  \includegraphics[width=5.2cm]{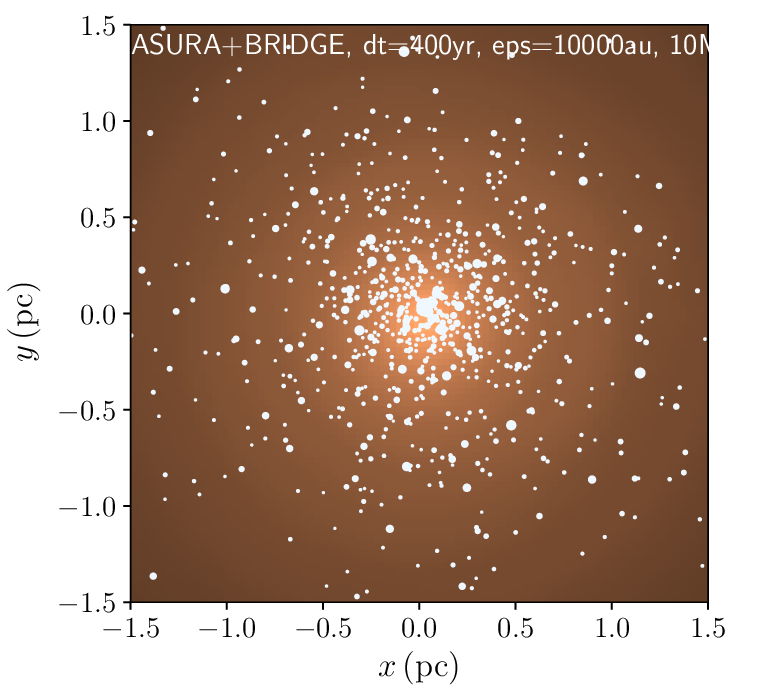}
 \end{center}
\caption{Gas and stellar distributions of initial condition (top left) and models AG3S3 (top middle), AG4S4 (top right), BG3S0T05 (bottom left), BG3S0T20 (bottom middle) and BG4S0T40 (bottom right) at 10\,Myr. The density plots are a slice through the mid-plane of the gas density. Points show the stars and the marker size depend on the stellar mass.}\label{fig:snap}
\end{figure*}

The gas sphere is modeled with $10^5$ particles.
We set the total gas mass to $844 M_{\odot}$, and the resulting individual gas-particle mass is $0.0844 M_{\odot}$. 
We assume that the gas is initially isothermal and its temperature is 10\,K according to \citet{2012MNRAS.420.1503P}.
We set the number of stellar particles to be $10^3$ and assign masses randomly sampled from the Salpeter mass function \citep{1955ApJ...121..161S} with the lower- and upper-mass cut-off of $0.1M_{\odot}$ and $100M_{\odot}$, respectively. However, the maximum stellar mass we obtained was $\sim 30M_{\odot}$ in our realization because of the random sampling from the mass function. The total stellar mass is $362M_{\odot}$. The initial condition is generated using \textsc{AMUSE} \citep{2013CoPhC.183..456P, 2013A&A...557A..84P,AMUSE}.
In the top left panel of Fig.~\ref{fig:snap}, we present the snapshot of the initial gas and stellar distribution.

\subsubsection{Parameters for integration}
In this test, we switch off cooling and heating processes and solve the evolution of gas adiabatically following \citet{2012MNRAS.420.1503P}. Here, the ratio of specific heat, $\gamma=5/3$, is adopted. 
We also switch off the star formation. 

In order to confirm the energy conservation and the gas and stellar distributions, we perform a series of simulations changing the Bridge timestep ($\Delta t_{\rm B}$) and the softening lengths for stars and gas ($\epsilon_{\rm s}$ and $\epsilon_{\rm g}$, respectively). Note that the softening length for star-gas interaction is also $\epsilon_{\rm g}$.
Two different sizes are used as the Plummer-type softening length, $\epsilon_{\rm g}$; they are 1000\,au and 10000\,au.
For $\epsilon_{\rm s}$, we test two cases: $\epsilon_{\rm s}=\epsilon_{\rm g}$ and $\epsilon_{\rm s}=0.0$.

The timestep for the Bridge scheme ($\Delta t_{\rm B}$) has to be chosen properly depending on the softening length and particle mass. We test some timesteps for different combinations of $\epsilon_{\rm g}$ and $\epsilon_{\rm s}$. In Table \ref{tb:EC}, we summarize the parameter sets of $\epsilon_{\rm g}$, $\epsilon_{\rm s}$, and $\Delta t_{\rm B}$ for each run.
We fix the other parameters such as the accuracy parameter which determines the individual timestep in the Hermite integrator used in \textsc{ASURA+BRIDGE}, $\eta = 0.1$ \citep{2008NewA...13..498N}. We use the sixth-order Hermite scheme for this test. 
We integrate the system up to 10\,Myr.

\begin{table}
  \caption{Parameters for embedded cluster models}{%
  \begin{tabular}{lcccc}
  \hline
  Name & $\epsilon_{\rm g}$ (au) & $\epsilon_{\rm s}$ (au) & $\Delta t_{\rm B}$ (yr) & code \\ 
  \hline
    AG3S3 & 1000 & 1000 & - & \textsc{ASURA} \\ 
    AG4S4 & 10000 & 10000 & - & \textsc{ASURA}\\ 
    BG3S0T20 & 1000 & 0 & 200 & \textsc{A+B} \\ 
    BG3S0T10 & 1000 & 0 & 100 & \textsc{A+B} \\ 
    BG3S0T05 & 1000 & 0 & 50 & \textsc{A+B} \\ 
    BGS3S3T20 & 1000 & 1000 & 200 & \textsc{A+B} \\ 
    BG4S0T80 & 10000 & 0 & 800 & \textsc{A+B} \\ 
    BG4S0T40 & 10000 & 0 & 400 & \textsc{A+B} \\ 
      \hline
    \end{tabular}}\label{tb:EC}
\begin{tabnote}
From left to right, columns represent the name of the models, the softening length for gas ($\epsilon_{\rm{g}}$) and stars ($\epsilon_{\rm s}$), the Bridge timestep ($\Delta t_{\rm B}$), and the code name (`A+B' is \textsc{ASURA+BRIDGE}), respectively.
The softening length between star and gas particles is equal to $\epsilon_{\rm g}$. 
\end{tabnote}
\end{table}

\subsubsection{Energy conservation}
\begin{figure}
 \begin{center}
  \includegraphics[width=8cm]{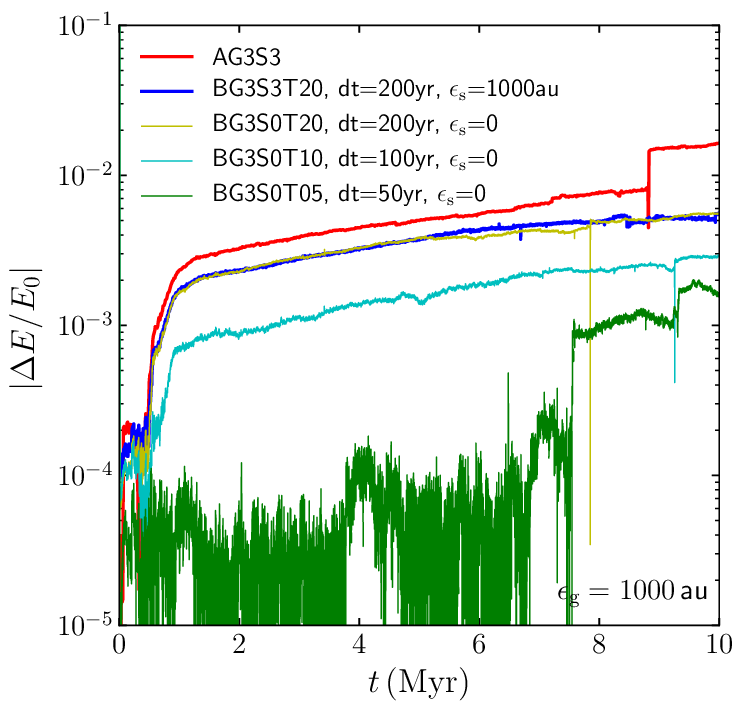} 
 \end{center}
\caption{Energy error as a function of time for models with $\epsilon _{g}=1000$\,au performed using \textsc{ASURA} (model AG3S3) and \textsc{ASURA+BRIDGE} with different Bridge timesteps (models BG3S3T20, BG3S0T20, BG3S0T10, and BG3S0T05).}\label{fig:Eerr_EC}
\end{figure}

First, we compare the energy conservation between \textsc{ASURA} and \textsc{ASURA+BRIDGE}. We here set the same softening length for gas and star particles for \textsc{ASURA+BRIDGE} ($\epsilon_{\rm s}=\epsilon_{\rm g}=1000$\,au; model BG3S3T20) and \textsc{ASURA} (model AG3S3). For the Bridge timestep, we adopt 200\,yr. 
In Fig.~\ref{fig:Eerr_EC}, we present the time evolution of the energy error for these models. Here, we count the kinetic and potential energy of stars and gas, the thermal energy of gas.
With $\epsilon_{\rm g}=1000$\,au, the energy error using \textsc{ASURA} (model AG3S3) monotonically increases and at around 9\,Myr, it suddenly jumps up. This is due to a close encounter of stars. Using \textsc{ASURA+BRIDGE} with the same setup (model BG3S3T20), the energy conserves better than \textsc{ASURA} and no energy jump occurs thanks to the Hermite integrator. The total energy conservation is better than \textsc{ASURA}. 

We next present the energy conservation of the cases without softening for stars. For model BG3S0T20, we set the softening length of stars to be zero, but adopt a parameter set the same as that of model BG3S3T20. The energy error of this model is shown in Fig.~\ref{fig:Eerr_EC}. The energy conservation of this model is similar to that of model BG3S3T20, and this result shows that the Hermite integrator can maintain the energy error sufficiently small even without softening length.

We also check the energy conservation with shorter Bridge timesteps (models BG3S3T10 and BG3S3T05 for $\Delta_{\rm t}=100$ and $50$, respectively). As the Bridge timestep decreases, the energy conservation improves, and with $\Delta t_{\rm B}=50$\,yr (model BG3S3T05), the energy error shows a random walk behavior, which is typical in integration with a tree scheme.

The energy conservation better than \textsc{ASURA} does not guarantee that the calculation with \textsc{ASURA+BRIDGE} is correct. 
In the top middle and bottom middle panels in Fig.~\ref{fig:snap}, we present the snapshots of models AG3S3 and BG3S0T20 at 10\,Myr, respectively.  We find that the cluster center in model BG3S0T20 shift from the center of gas distribution. This is caused by the too large Bridge timestep.
The bottom left panel shows the snapshot for model BG3S0T05, in which the Bridge timestep is a quarter of model BG3S0T20. In this model, we do not find the shift of the stellar distribution. We will discuss the choice of Bridge timesteps in the last paragraph of this section.

\begin{figure}
 \begin{center}
  \includegraphics[width=8cm]{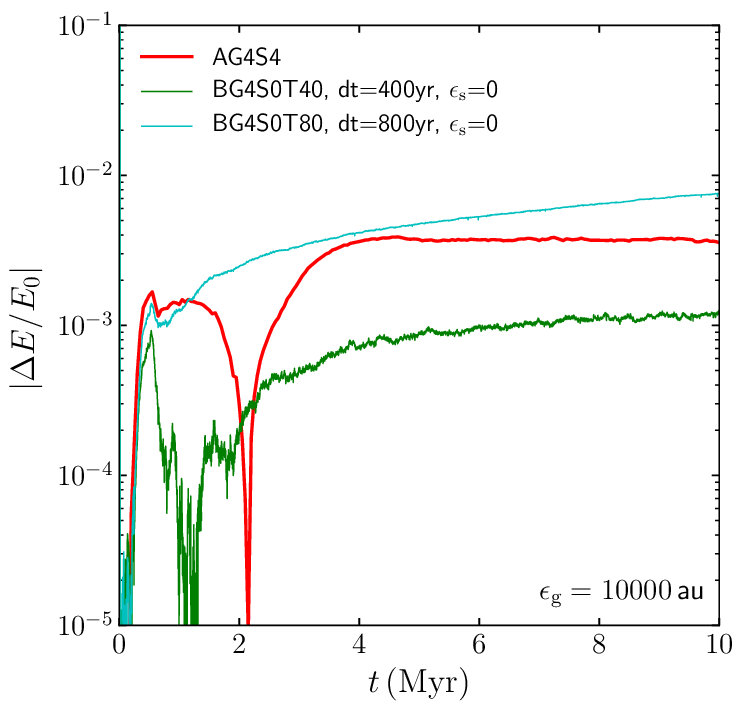} 
 \end{center}
\caption{Same as Fig.~\ref{fig:Eerr_EC} but for models with $\epsilon _{g}=10000$\,au performed using \textsc{ASURA} (model AG4S4) and \textsc{ASURA+BRIDGE} with different Bridge timesteps (models BG4S0T40 and BG4S0T80)}.\label{fig:Eerr_EC_soft}
\end{figure}

With a larger softening length for gas, the energy conservation is better. The energy conservation with $\epsilon_{\rm s}=\epsilon_{\rm g}=10000$\,au using \textsc{ASURA} (model AG4S4) is shown in Fig.~\ref{fig:Eerr_EC_soft}. The snapshot at 10\,Myr is shown in the top right panel of Fig.~\ref{fig:snap}. The energy conservation of model AG4S4 is better than that of AG3S3 and does not evolve linearly with time, thanks to the larger softening length. For comparison between \textsc{ASURA} and \textsc{ASURA+BRIDGE}, we also show the snapshot of model BG4S0T40 at $t=10$\,Myr in the bottom right panel of Fig.~\ref{fig:snap}. The energy error for models BG4S0T40 and BG4S0T80 are presented in Fig.~\ref{fig:Eerr_EC_soft}.
The energy conservation of \textsc{ASURA+BRIDGE} is better than \textsc{ASURA}, if we take a sufficiently small Bridge timestep.

\begin{figure*}
 \begin{center}
  \includegraphics[width=8cm]{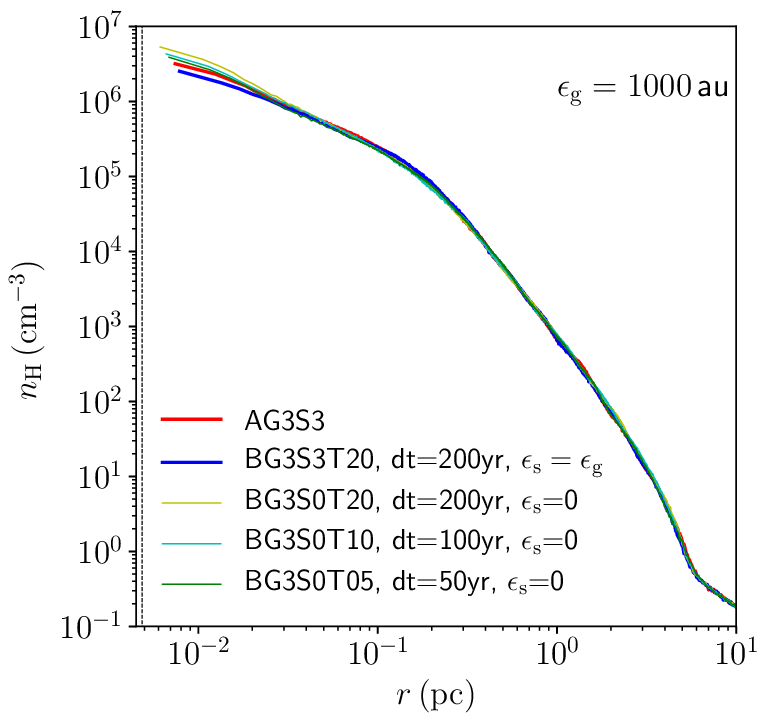} \includegraphics[width=8cm]{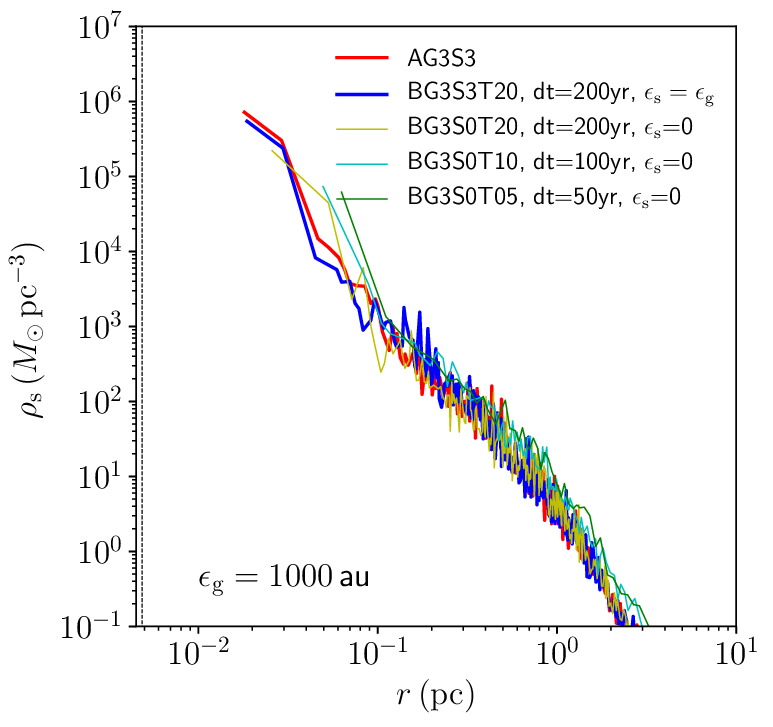} \\
  \includegraphics[width=8cm]{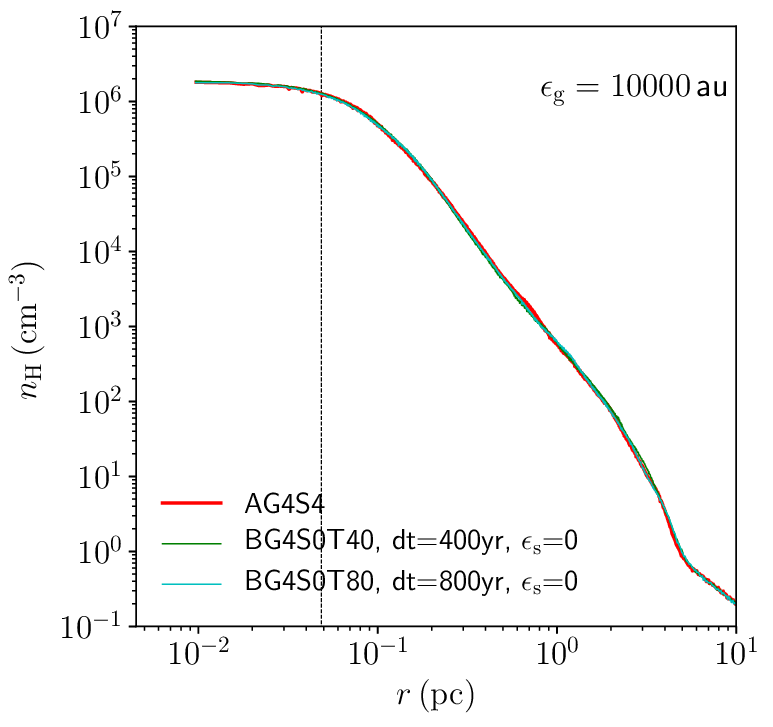}
  \includegraphics[width=8cm]{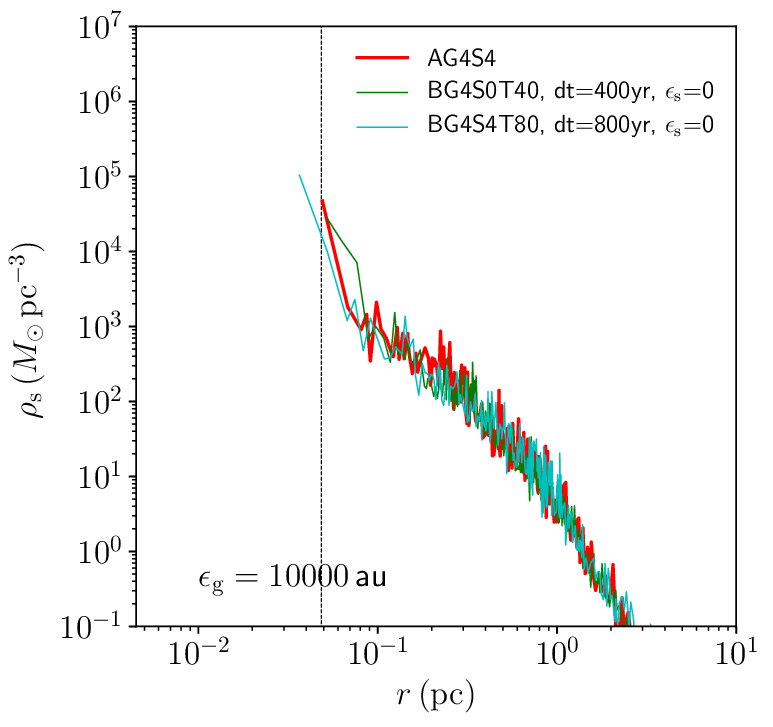} \\
 \end{center}
\caption{Density profiles of gas (left) and star (right) at 10\,Myr using \textsc{ASURA} and \textsc{ASURA+BRIDGE}. Top panels are for $\epsilon_{\rm g}=1000$\,au and bottom panels for $\epsilon_{\rm g}=10000$\,au. The dotted line indicates the softening length for gas. }\label{fig:prof_gas}
\end{figure*}

We estimate Bridge timesteps ($\Delta t_{\rm B}$) necessary for these models by assuming that the timestep must be smaller than $\sim 1/100$ of the free-fall time of gas. In the left panel of Fig.~\ref{fig:prof_gas}, we present the density profile of gas and stars at the end of the simulation (10\,Myr) for the runs with $\epsilon_{\rm g}=10000$ and 1000\,au, respectively.
For $\epsilon_{\rm g}=10000$\,au, the highest gas density (number density of hydrogen) is $\sim 1\times10^{6}$\,cm$^{-3}$, and the free-fall time for this density is $\sim 5\times 10^{4}$\,yr. Therefore, the necessary timestep is estimated to be $\sim 500$\,yr. 
For $\epsilon_{\rm g}=1000$\,au, the gas density reaches $\sim 1\times 10^{7}$\,cm$^{-3}$. The corresponding free-fall time is $\sim 16000$\,yr, and the necessary timestep is estimated to be $\sim 160$\,yr. 
These estimates are roughly consistent with the energy conservation shown in Figs.~\ref{fig:Eerr_EC} and \ref{fig:Eerr_EC_soft}. As is shown in the bottom middle panel of Fig.~\ref{fig:snap} (model BG3S0T20), a too large Bridge timestep causes a drift of stellar systems with respect to the gas distribution. 

\subsubsection{Structures of embedded clusters}
We compare the structures of gas and stellar distributions with \textsc{ASURA} and \textsc{ASURA+BRIDGE}. In Fig.~\ref{fig:prof_gas}, we present the density profiles of gas and stars at 10\,Myr for all models.
The stellar density profiles are drawn using a method presented in \citet{1985ApJ...298...80C}. The density center of the stellar distribution is also determined using the method of \citet{1985ApJ...298...80C}, while the center of the gas distribution is defined by the position of the densest gas particle. The data points are smoothed every 10 points. 

With $\epsilon_{\rm g}=10000$\,au, we find a core structure of gas due to the softened gravity. In the case with $\epsilon_{\rm g}=1000$\,au, the gas density continuously increases toward the center. We confirm that the distribution of gas is identical for \textsc{ASURA+BRIDGE} and \textsc{ASURA}. 

The stellar distributions in the outer regions of the clusters are also consistent for both codes. With a softening length for star particles, however, we find an artificial core in the cluster center, which is clear in the case of $\epsilon_{\rm s}=1000$\,au (see red and blue curves in Fig.~\ref{fig:prof_gas}). This is seen in both \textsc{ASURA} and \textsc{ASURA+BRIDGE}, if we set a softening length for stars. 
Without softening, stars can scatter each other, and as a consequence, the artificial clump disappears. With $\epsilon_{\rm s}=10000$\,au, the core size of the gas distribution is similar to the softening length, and therefore stellar distribution is not concentrated to the cluster center for both models with/without softening.
Thus, simulations without gravitational softening are important to follow the structure and the dynamical evolution of star clusters.

\subsection{Test 2: Star cluster formation model}
As the second test, we perform a series of simulations of star cluster formation starting from turbulent molecular clouds that have an initial mass and size the same as a model used in \citet{2003MNRAS.343..413B}. We here include star formation and investigate the star formation history and the structures of the formed clusters.

\subsubsection{Initial conditions: Turbulent molecular clouds}
As an initial condition, we adopt a homogeneous spherical molecular cloud with a turbulent velocity field of $k=-4$ following \citet{2003MNRAS.343..413B}. 
The total gas mass ($M_{\rm g}$) and initial radius ($R_{g}$) of the molecular cloud are $1000 M_{\odot}$ and $0.5$\,pc, respectively. As a consequence, the initial density and the free-fall time ($t_{\rm ff}$) are $7.6\times 10^{-4}$\,cm$^{-3}$ and $0.18$\,Myr, respectively. The ratio between the kinetic energy ($|E_{\rm k}|$) to the potential energy ($|E_{\rm p}|$) is unity. We generate the initial condition using \textsc{AMUSE} \citep{2013CoPhC.183..456P, 2013A&A...557A..84P,AMUSE}. The initial gas temperature is set to be 20\,K, and the equation of state of the gas follows $P=(\gamma-1) \rho u$, where $P$ is the pressure, $\rho$ is the density, and $u$ is the specific internal energy.
We hereafter call this model as `star cluster formation model'. 

In order to understand the dependence of the results on the resolution, we adopt three different mass resolutions of gas particles ($m_{\rm g}$) as $m_{\rm g}=0.002$ (High), $0.01$ (Middle), and $0.1\,M_{\odot}$ (Low), where $m_{\rm g}=0.002\,M_{\odot}$ is the same as that of \citet{2003MNRAS.343..413B}.
We set the softening length of gas ($\epsilon_{\rm g}$) depending on the mass resolution; $\epsilon_{\rm g}=2800$, $7000$, and $14000$\,au for models High, Middle, and Low. In this test, we do not use softening for stellar particles. 

\begin{table*}
\caption{Models and results for star cluster formation.}
  \begin{tabular}{lccccccccccc}
      \hline
      Name & $N_{\rm g}$ & $m_{\rm g}$ & $\epsilon_{\rm g}$ & $n_{\rm th}$ & $c_{\star}$ & $r_{\rm max}$ & $\Delta t_{\rm B}$ & $M_{\rm s, 0.5Myr}$ & $N_{\rm run}$ & $\langle M_{\rm s, 1Myr}\rangle$ \\
       & & $(M_{\odot})$ & (au) & (cm$^{-3}$) & & (pc) &  (yr) &$(M_{\odot})$ & & $(M_{\odot})$ \\
      \hline
      High  & $5\times 10^5$ & $0.002$ & $2.8\times 10^3$ & $1.8\times 10^6$ & $0.02$ & $0.05$ & $10$ & $443$ & $3$ & $599\pm 129$\\
      
      Middle & $10^5$ & $0.01$ & $7\times 10^3$  & $1.8\times10^6$ & $0.02$ & $0.1$ & $50$ & $272$ & $5$ & $456\pm 119$\\
      
      Middle-c01 & $10^5$ & $0.01$ & $7\times 10^3$  & $1.8\times10^6$ & $0.1$ & $0.1$ & $50$ & $383$ & $1$ & -\\
      
      Middle-l & $10^5$ & $0.01$ & $1.4\times 10^4$ & $7.2\times10^4$ & $0.02$ & $0.15$ & $100$ & $181$  & $1$ & -\\
      
      Low & $10^4$ & $0.1$ & $1.4\times 10^4$ & $5.0\times10^5$ & $0.02$ & $0.15$ & $100$ & $154$ & $10$ & $360\pm 119$\\
      
      Low-c01 & $10^4$ & $0.1$ & $1.4\times 10^4$ & $5.0\times10^5$ & $0.1$ & $0.15$ & $100$ & $355$  & $1$ & -\\
      
      Low-l & $10^4$ & $0.1$ & $7\times 10^4$ & $3.0\times10^3$ & $0.02$ & $0.15$ & $100$ & $187$ & $1$ & -\\
      \hline
    \end{tabular}\label{tb:IC_B}
\begin{tabnote}
From the left: model name, the number of gas particles ($N_{\rm{g}}$), gas particle mass ($m_{\rm g}$), softening length for gas ($\epsilon_{\rm g}$), star formation threshold density ($n_{\rm th}$), star formation efficiency ($c_{*}$), the maximum search radius ($r_{\rm max}$), timestep for Bridge ($\Delta t_{\rm B}$), stellar mass at 0.5\,Myr for seed 1 ($M_{\rm s, 0.5Myr}$), the number of runs with different random seeds ($N_{\rm run}$), the average mass at 0.5\,Myr, if multiple runs are preformed ($\langle M_{\rm s, 0.5Myr}\rangle$).
\end{tabnote}
\end{table*}

\subsubsection{Integration and star formation}
We integrate all models using \textsc{ASURA+BRIDGE} with the sixth-order Hermite integrator. Star formation is also assumed using the model described in section 2.3 \citep[see also ][]{2020arXiv200512906H}.  

We first briefly discuss our choice of the softening lengths. The softening length for gas ($\epsilon_{\rm g}$) and threshold density for star formation ($n_{\rm th}$) should depend on the mass resolution. If we assume that the Jeans mass is resolved by 64 SPH particles, using the sound speed for our minimum temperature (20\,K), we can estimate the density which we can resolve (i.e., $n_{\rm th}$) and the corresponding Jeans length (i.e., $\epsilon_{\rm g}$). 
For model High ($m_{\rm g}=0.002$\,$M_{\odot}$), we obtain $\epsilon_{\rm g}=2.8\times 10^3$\,au (0.014\,pc) and $n_{\rm th}=1.8\times 10^6$\,cm$^{-3}$.
The softening length should be larger and the threshold density should be lower for lower mass resolution of gas. Although we can adopt those from the Jeans mass resolution, we chose the softening length and threshold density for models Middle and Low to reproduce the results of model High. We set $\epsilon_{\rm g}=7.0\times10^3$\,au ($=0.035$\,pc) and $n_{\rm th}=7.2\times 10^4$\,cm$^{-3}$ for model Middle, and $\epsilon_{\rm g}=1.4\times 10^4$\,au ($=0.07$\,pc) and $n_{\rm th}=5.0\times 10^{5}$\,cm$^{-3}$ for model Low. These softening lengths are smaller than the Jeans length, and the threshold densities are higher than that for the Jeans length, but close to the observed density of star-forming cores which is typically $10^4$--$10^5$\,cm$^{-3}$ \citep{2002ApJ...575..950O, 2007ApJ...665.1194I, 2015ApJS..217....7S}.

In order to understand the dependence of the results on these parameters, we also test a model the same as model Low, but with a lower spatial resolution ($\epsilon_{\rm g}=7\times 10^4$\,au $=0.35$\,pc) and a lower star formation threshold density ($n_{\rm th}=3.0\times 10^3$\,cm$^{-3}$). We call this model as model Low-l. These values correspond to the Jeans length and threshold density assuming 50 SPH particles for the Jeans mass. 

The mass of forming stars is determined by the initial mass function we assume and local gas mass determined by the maximum search radius ($r_{\rm max}$) for star formation.
We adopt the Kroupa mass function with an upper and lower mass limit of $0.1$ and $150\,M_{\odot}$.
For $r_{\rm max}$, we chose the value depending on the mass resolution and softening length. The value of $r_{\rm max}$ should be sufficiently small. If it is too small, however, the mass of the forming stars is limited due to $r_{\rm max}$ because the mass of the forming star is limited to the half of gas mass within $r_{\rm max}$. By calculating the mass within $r_{\rm max}$ with a given $n_{\rm th}$, we adopt $r_{\rm max}$ to form a $\sim 100\,M_{\odot}$ star; $r_{\rm max}=0.05$, 0.1, and 0.15\,pc for models High, Middle, and Low. These parameters are also discussed in \citet{2020arXiv200512906H}. The threshold temperature for star formation is set to be 20\,K for all models. Since we set the minimum temperature of the gas to be $20$\,K, gas with 20\,K can form stars. These parameters are summarized in Table \ref{tb:IC_B}. 

Star formation efficiency per free-fall time ($c_{\star}$) is another parameter that we have to choose for star formation. 
We adopt $c_{\star}=0.02$ for our fiducial model. According to previous study on galaxy simulations, the value of $c_{\star}$ should not largely change the star formation rate, if we set $n_{\rm th}$ large enough \citep{2008PASJ...60..667S}. For models Middle and Low, we also perform runs with larger values of $c_{*}$; $c_{*}=0.1$ for models Low and Middle (models Low-c01 and Middle-c01, respectively) in order to understand the dependence on $c_{\star}$. 
All the models are summarized in table \ref{tb:IC_B}.

We adopt the Bridge timestep ($\Delta t_{\rm B}$) of 10\,yr for model High and 100\,yr for the others. These are sufficiently small for energy conservation. The accuracy parameter for the Hermite integrator is set to be 0.2 \citep{1992PASJ...44..141M,2008NewA...13..498N}.
We continued the simulations up to 1\,Myr, which is $\sim 4 t_{\rm ff}$, for all models. For some models, we continue the simulation up to 1.2\,Myr. 

We performed 3, 5, and 10 runs for models High, Middle, and Low using different random seeds of the initial turbulent velocity field in order to see the run-to-run variation. The number of runs $N_{\rm run}$ for each model is also summarized in Table \ref{tb:IC_B}.

\begin{figure*}
 \begin{center}
\includegraphics[width=5.3cm]{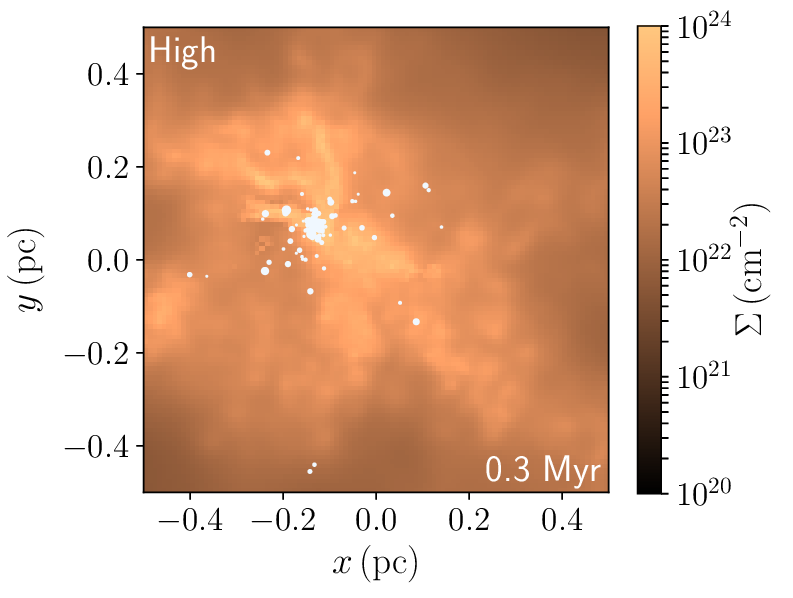}
\includegraphics[width=5.3cm]{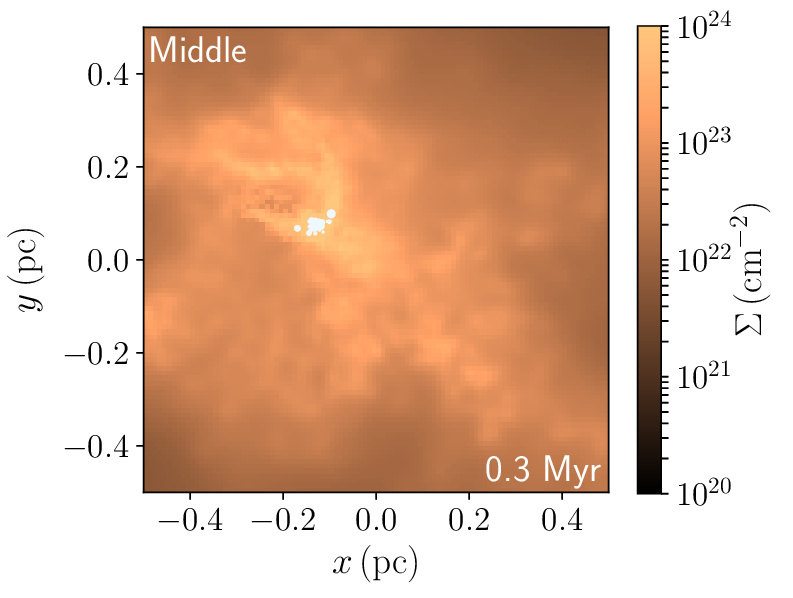}
 \includegraphics[width=5.3cm]{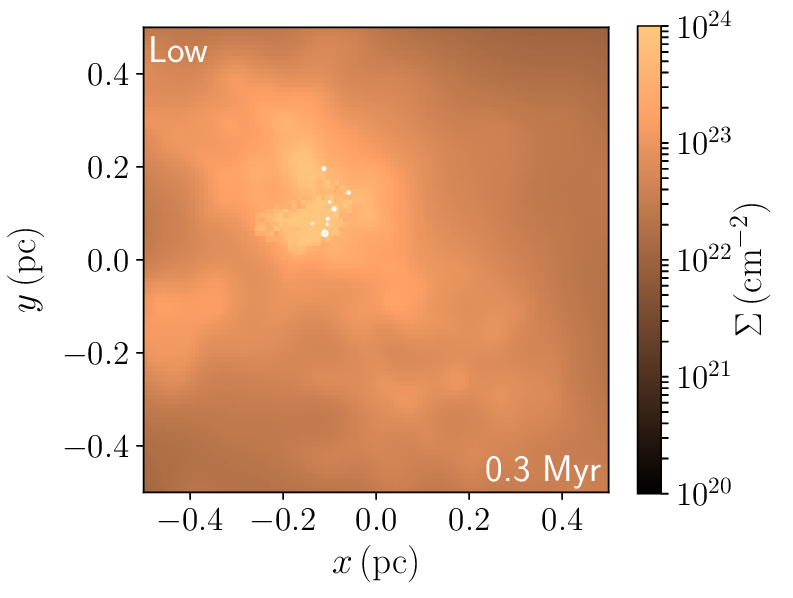}\\
 
 \includegraphics[width=5.2cm]{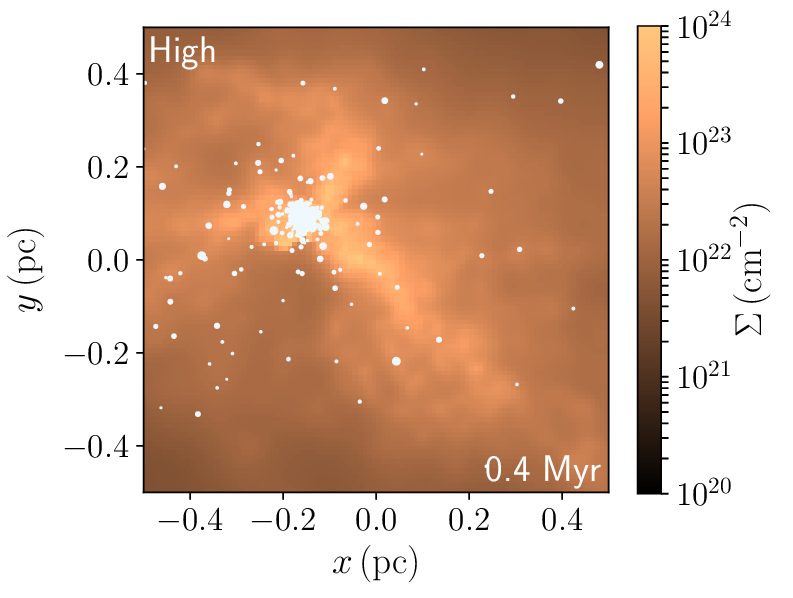}
  \includegraphics[width=5.2cm]{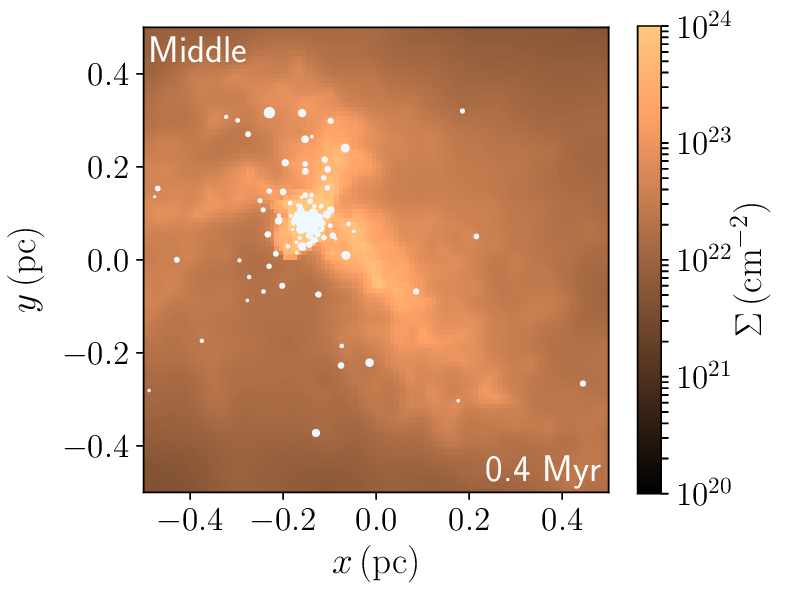}
 \includegraphics[width=5.2cm]{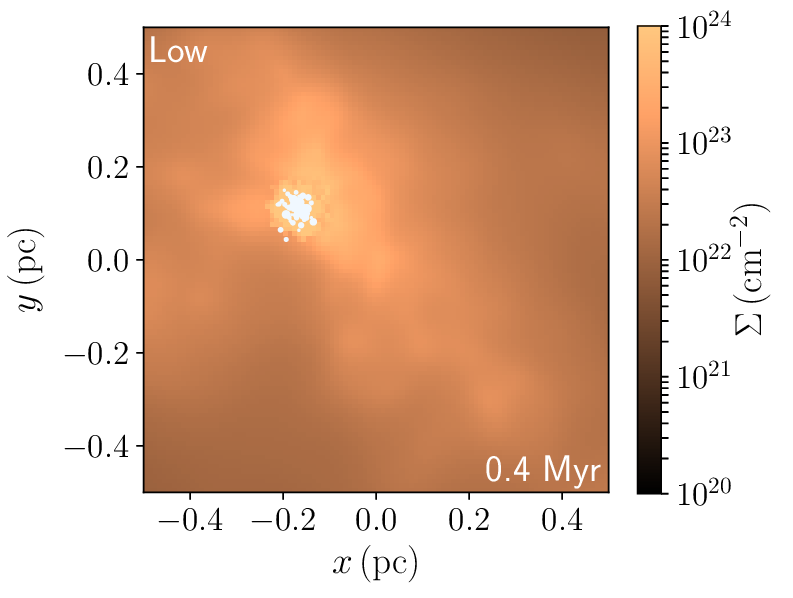}\\
 
 \includegraphics[width=5.2cm]{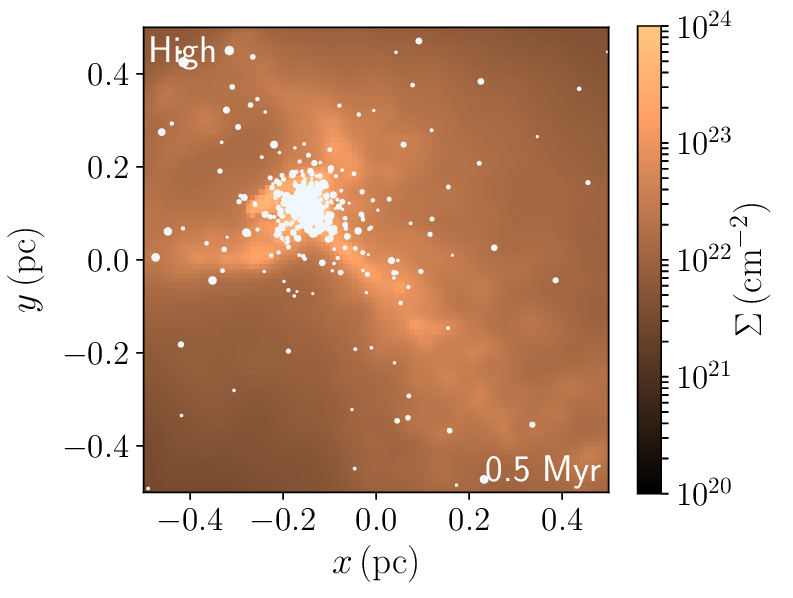}
 \includegraphics[width=5.2cm]{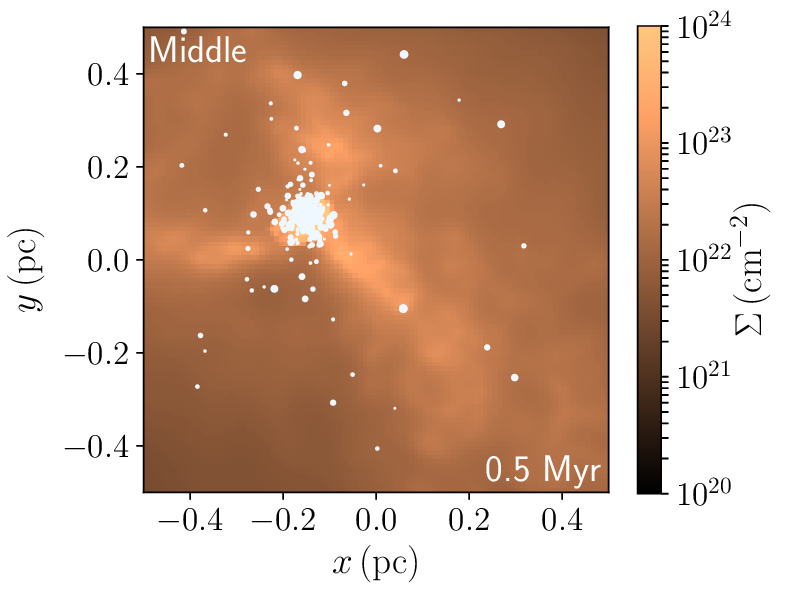}
 \includegraphics[width=5.2cm]{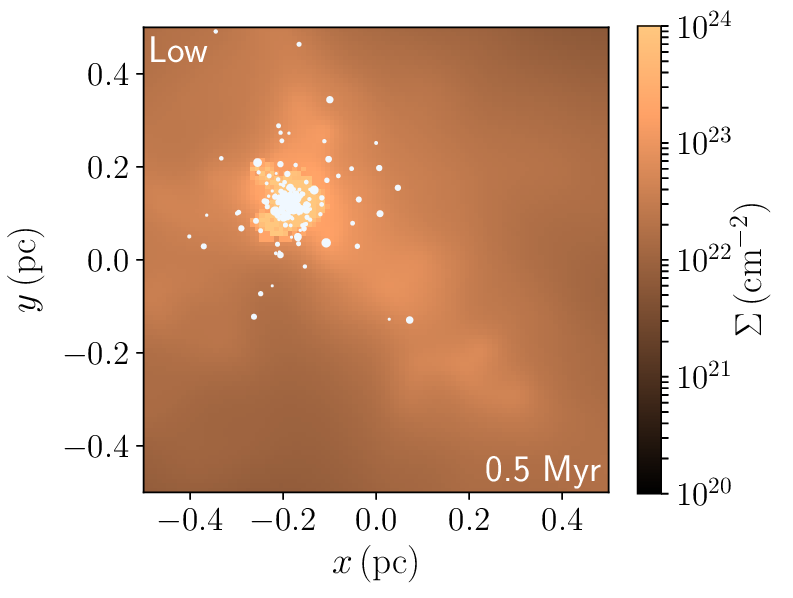}\\
 
 \includegraphics[width=5.2cm]{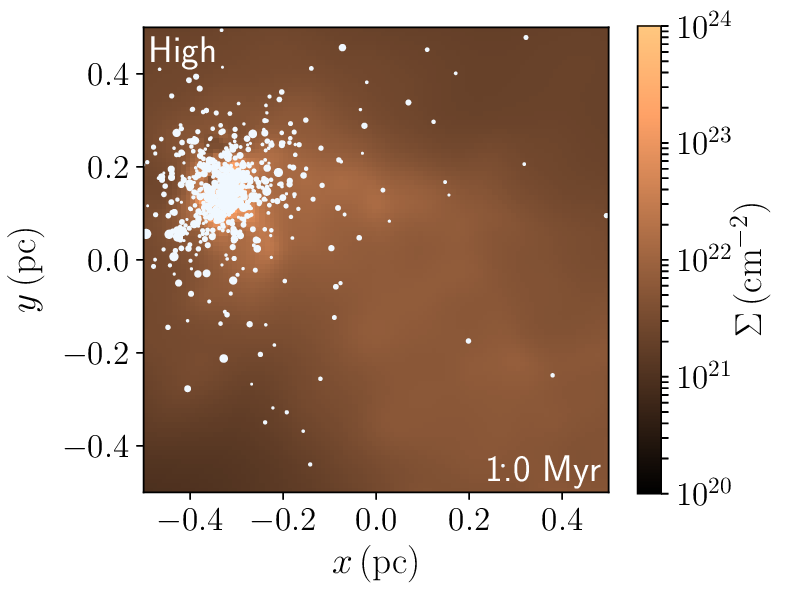}
 \includegraphics[width=5.2cm]{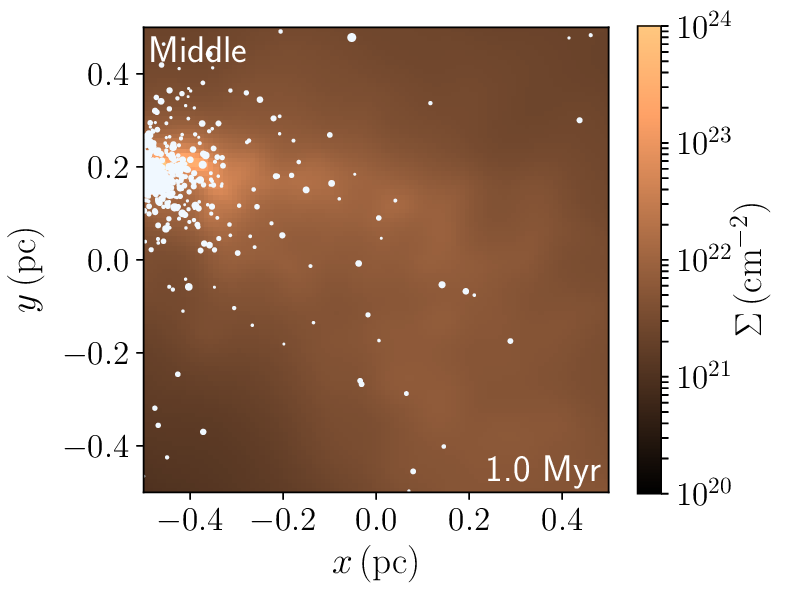}
 \includegraphics[width=5.2cm]{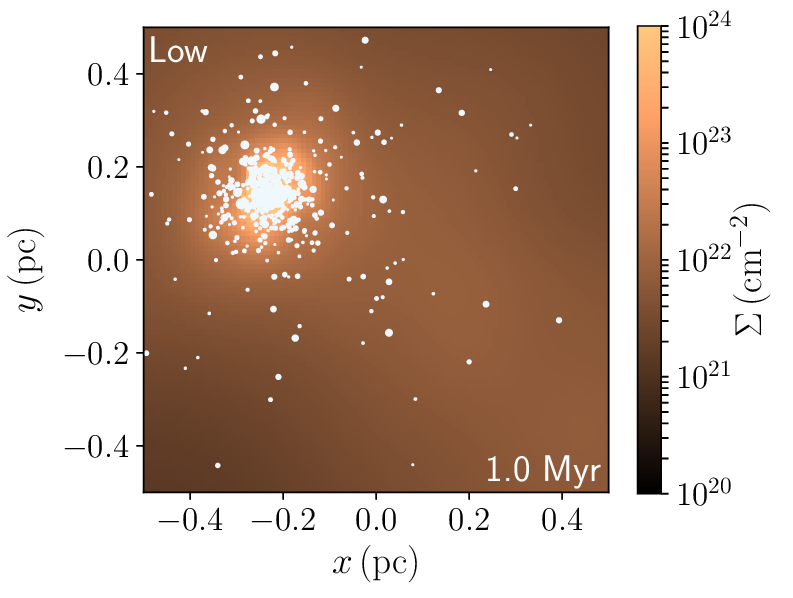}\\
  
 \end{center}
\caption{Snapshots of star cluster formation models for Models High, Middle, and Low from left to right. White dots shows stars, and the sizes indicate the stellar masses. Color contour shows the surface density of the gas.}\label{fig:snap_Bonnell}
\end{figure*}

\subsubsection{Evolution of the total stellar mass}
\begin{figure}
 \begin{center}
  \includegraphics[width=7.8cm]{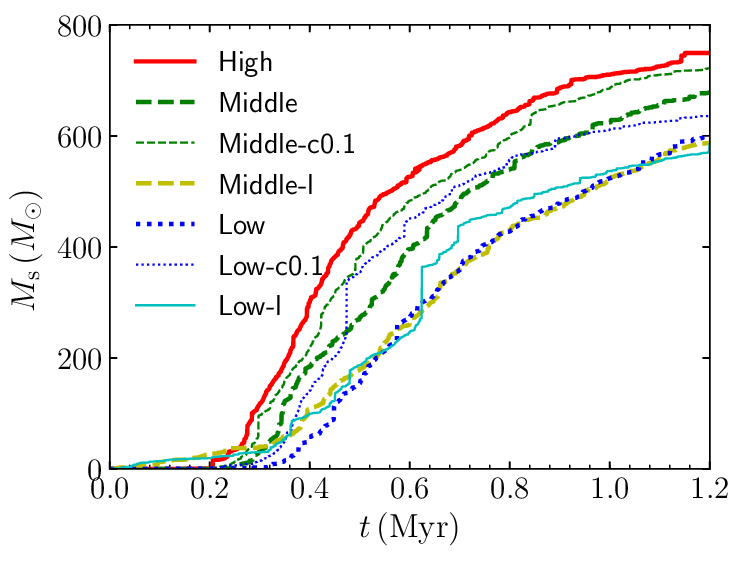} 
 \end{center}
\caption{Time evolution of the total stellar mass for the same random seed for the initial turbulent velocity field of the molecular cloud for star cluster formation model.}\label{fig:Bonnell_s2}
\end{figure}

In Fig.~\ref{fig:snap_Bonnell}, we present the snapshots of models High, Middle, and Low, among which the random seed for the initial turbulent velocity field is the same. The evolution of their total stellar masses is shown in Fig.~\ref{fig:Bonnell_s2}. The total stellar mass of model High at $t=0.5$\,Myr is 443\,$M_{\odot}$, which is similar to that in \citet{2003MNRAS.343..413B}  ($\sim 580\,M_{\odot}$ at 0.52\,Myr). Comparing models with different mass resolution but the same random seed, the total stellar masses at 0.5\,Myr are 272 and 154\,$M_{\odot}$ for models Middle and Low, respectively. Thus, the total stellar mass decreases as the resolution decreases. We continue the simulation until $t=1.2$\,Myr, at which the star formation almost ends. The difference in the total stellar mass among different mass resolution models becomes smaller; $750$, $678$, and $599$\,$M_{\odot}$ for models High, Middle, and Low. 

Even if we increase the star formation efficiency ($c_{\star}$) an order of magnitude larger (see models Middle-c01 and Low-c01), the final stellar mass increases only $\sim 10$\,\%. This result is consistent with that obtained in galaxy simulations \citep{2008PASJ...60..667S} because the amount of dense gas that satisfies the star formation condition is limited and the local free-fall time of such gas is much shorter than the simulating timescale. 

\begin{figure}
 \begin{center}
  \includegraphics[width=7.8cm]{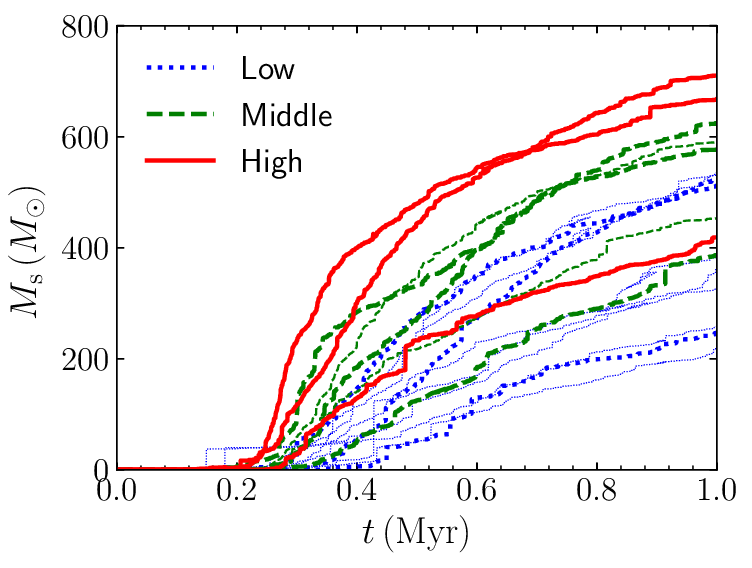} 
 \end{center}
\caption{Run-to-run variation in stellar mass evolution for star cluster formation models (model Low, Middle, and High). Thick curves in models Middle and Low indicate the three models which have the three random seeds for the turbulence used in model High. }\label{fig:Bonnell_m01}
\end{figure}

The random seeds for the initial turbulent velocity fields make a large difference in the distribution of stars and final stellar mass. In order to investigate such run-to-run variations, we perform three runs for model High with different random seeds for the initial turbulent velocity field up to 1\,Myr. We show the time evolution of the total stellar mass of these runs in Fig.~\ref{fig:Bonnell_m01}. As shown in this figure, one of them forms only $\sim 400\,M_{\odot}$ stars in total by 1\,Myr, although the others form more than $600\,M_{\odot}$ stars. We in addition perform 5 and 10 runs for models Middle and Low with different random seeds. The maximum mass is twice as large as the minimum one. Irrespective of the mass resolution, models with the random seed which results in the formation of less number of stars always form stars less than those with the other seeds. Thus, the forming stellar mass slightly depends on the mass resolution, but we can reproduce the variation depending on the turbulent velocity fields. We summarize the averaged total masses at 1\,Myr with standard deviations in table~\ref{tb:IC_B}. 

We also show the results with a larger softening length and lower threshold density for star formation (models Middle-l and Low-l). In these models, star formation starts in earlier time (see Fig.~\ref{fig:Bonnell_s2}). However, the final stellar mass does not change much from those with a smaller softening length especially in the case of the lowest mass resolution.

\subsubsection{Structures of formed star clusters}
\begin{figure}
 \begin{center}
  \includegraphics[width=7.8cm]{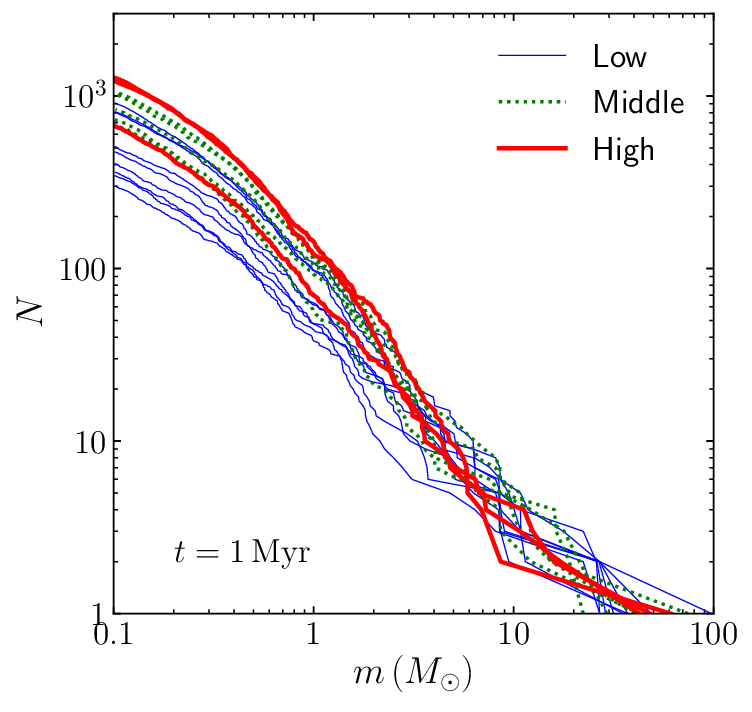} 
 \end{center}
\caption{Run-to-run variations of stellar mass function at 1\,Myr for star cluster formation models  (model Low, Middle, and High).}\label{fig:Bonnell_m01_MF}
\end{figure}

We investigate the structures of the formed star clusters in our simulations and the dependence on the resolution. 
In Fig.~\ref{fig:Bonnell_m01_MF}, we present the stellar mass functions obtained from our simulations at 1\,Myr. We confirm that formed stars follow the Kroupa mass function we gave. Although we set the maximum mass of the mass function to be $100M_{\odot}$, we did not find such a massive star in this simulation set. This is simply due to the probability to form such a massive star in $\sim 500M_{\odot}$ star cluster is too low. 

\begin{figure}
 \begin{center}
  \includegraphics[width=7.8cm]{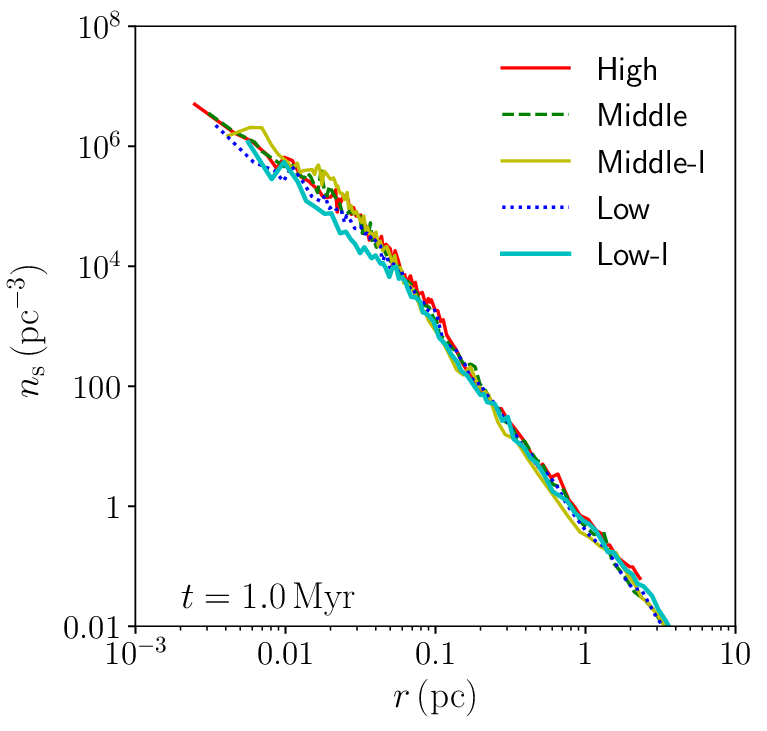}
 \end{center}
\caption{Density profiles of the formed clusters at 1\,Myr for cluster formation models with the same random seed for the initial turbulence but different mass resolution and softening length for gas. }\label{fig:density_prof_B_s2}
\end{figure}

We compare the density profiles of formed clusters.
In the density profiles, we find a small difference in the resolution. In Fig.~\ref{fig:density_prof_B_s2}, we present the density profiles of the runs with the same random seed, but different mass resolution and softening lengths. In the lowest resolution with a larger softening length (model Low-l), the formed cluster has a density slightly lower compared with the others. This is due to the large softening length of gas, although we did not assume gravitational softening for stars. For the other models, the density profiles of the formed clusters are identical.

\begin{figure}
 \begin{center}
  \includegraphics[width=7.8cm]{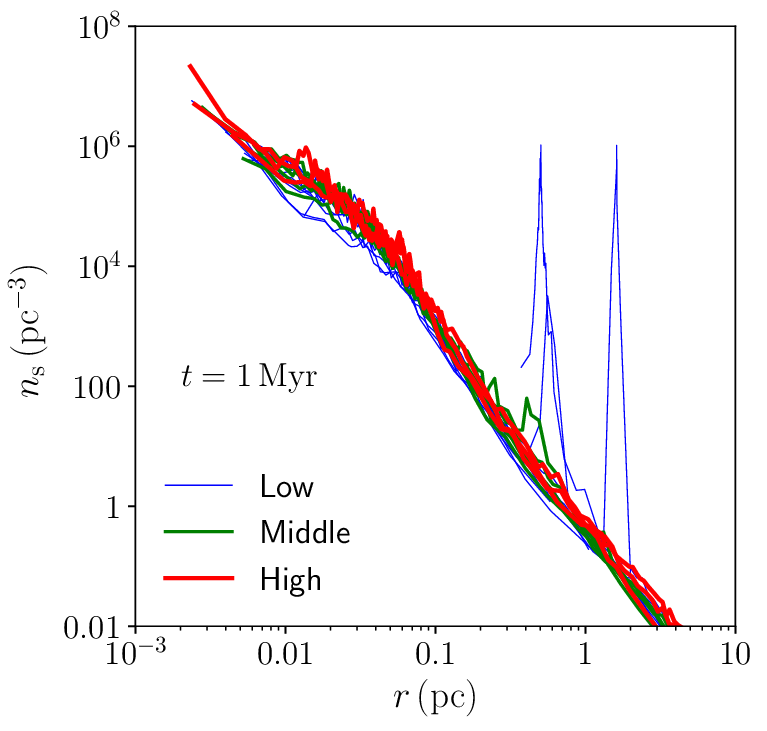}
 \end{center}
\caption{Run-to-run variations of density profiles of stars at 1\,Myr for cluster formation models (model Low, Middle, and High). Thick and thin curves are the same as Fig.~\ref{fig:Bonnell_m01}.}\label{fig:density_prof_B_10}
\end{figure}

\begin{figure}
 \begin{center}
  \includegraphics[width=7.8cm]{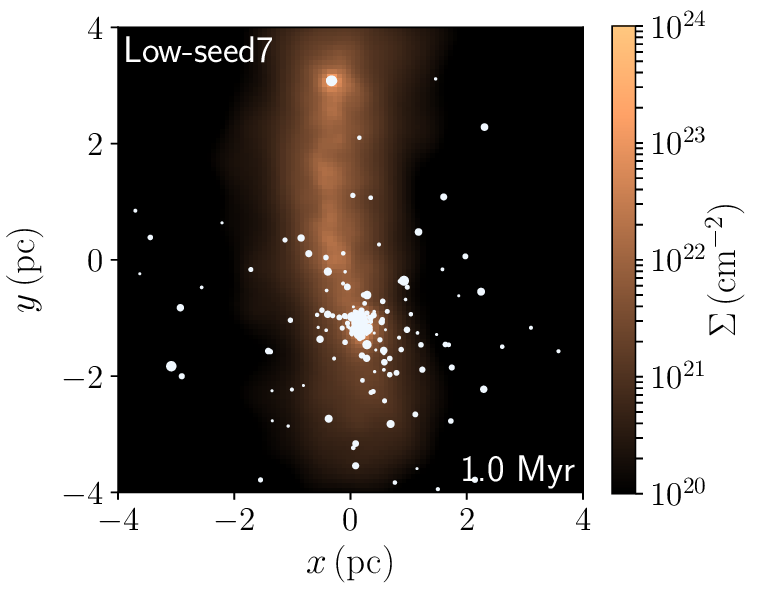}
 \end{center}
\caption{Snapshot at 1\,Myr for model Low with seed 7. The dots and colors are the same as Fig.~\ref{fig:snap_Bonnell}.}\label{fig:snapshot_s7}
\end{figure}

We also investigate the run-to-run variations in the density profiles of formed star clusters. 
In Fig.~\ref{fig:density_prof_B_10}, we present the density profiles with different random seeds for the initial turbulence for models Low, Middle, and High. For some runs, we find sub-clusters, which are shown as spikes in the density profiles. In Fig.~\ref{fig:snapshot_s7}, we show a snapshot of a model that includes a sub-cluster (Model Low, seed 7). Since the clusters form via the merger of smaller clumps, we sometimes find a surviving sub-cluster at 1\,Myr.

\subsection{Test 3: Star cluster complex model}
We in addition present the results of a larger-scale simulation, in which multiple star clusters, so-called star cluster complex, form in a turbulent molecular cloud. 

\subsubsection{Initial condition and integration}
We adopt an initial condition similar to \citet{2016ApJ...817....4F}, which is also a homogeneous spherical molecular cloud with turbulence similar to the model we adopted in the previous subsection but more massive. We adopt the initial radius and mass of the cloud as 10\,pc and $4\times 10^5\,M_{\odot}$, respectively. We set the gas-particle mass to be $0.1 M_{\odot}$. With this set-up, the initial density of the cloud is $3.8\times 10^3$\,cm$^{-3}$ and the initial free-fall time is 0.83\,Myr. \citep[See also ][]{2015PASJ...67...59F, 2016ApJ...817....4F}. The initial gas temperature is set to be 20\,K, and the equation of state of the gas follows $P=(\gamma-1) \rho u$.
We adopt parameters the same as those for model Low, but we adopt 200\,yr for $\Delta t_{\rm B}$. These parameters are summarized in Table~\ref{tb:IC_F}.

We first integrate the system using the Hermite scheme, but we switch the integrator to \textsc{PeTar} at 0.75\,Myr. At this time, the number of stellar particles was $\sim 8000$. We set the outer cutoff radius for the P$^3$T scheme ($r_{\rm out}$) to be $10^{-3}$\,pc. The inner cutoff radius is $0.1 r_{\rm out}$, which is the default value of \textsc{PeTar}. The slow-down factor for SDAR is also the default value ($10^{-4}$). During one Bridge step ($\Delta t_{\rm B}$), \textsc{PeTar} integrates 1024 tree steps, i.e., $\Delta t_{\rm soft} = \Delta t_{\rm B}/1024$. This may sound very small, but with Hermite scheme, the minimum timestep sometimes reaches $4\times 10^{-6} \Delta t_{\rm B} \sim 0.3$\,day. The integration time per $\Delta t_{\rm B}$ for $\sim 10^4$ stars with the Hermite scheme was $\sim 60$\,second using 80 CPU cores on Cray XC50, on the other hand, that with \textsc{PeTar} was $\sim 10$ second. Since the best performance of \textsc{PeTar} is obtained when the particle number per core is more than 1000 \citep{Wang2020c}, the performance of \textsc{PeTar} compared with Hermite scheme would be better as the number of stars increases.

\begin{table*}
\caption{Model for star cluster complex formation}
\begin{tabular}{lcccccccccccccc}
\hline
   Name  & $M_{\rm g}$  & $m_{\rm g}$ & $R_{\rm g}$& $n_{\rm ini}$ & $t_{\rm ff,ini}$  & $\alpha_{\rm vir}$ & $\epsilon_{\rm g}$ & $\epsilon_{\rm s}$ & $n_{\rm th}$ & $c_{\star}$ & $r_{\rm max}$ & $\Delta t_{\rm B}$ & $\Delta t_{\rm soft}$ & $r_{\rm out}$ \\
       & $(M_{\odot})$ & $(M_{\odot})$ & (pc) &  (cm$^{-3}$) & (Myr) &  & (pc) & (pc) & (cm$^{-3}$) &  & (pc) & (yr) & & (pc)\\ 
      \hline
  F16 & $4\times 10^5$ & $0.1$ & 10 & $3800$ & $0.83$ & $1$  & $0.07$ & $0.0$ & $5\times 10^5$ & 0.02 & $0.2$ & $200$ & $\Delta t_{\rm B}/1024$ & $10^{-3}$ \\
      \hline
    \end{tabular}\label{tb:IC_F}

\begin{tabnote}
From the left: model name, initial cloud mass ($M_{\rm g}$), gas-particle mass ($m_{\rm g}$), initial cloud radius ($R_{\rm g}$), initial cloud density ($n_{\rm ini}$), initial free-fall time ($t_{\rm ff, ini}$), initial virial ratio ($\alpha_{\rm vir}$), softening length for gas ($\epsilon_{\rm g}$) and stars ($\epsilon_{\rm s}$), star formation threshold density ($n_{\rm th}$), the maximum search radius ($r_{\rm max}$), timestep for Bridge ($\Delta t_{\rm B}$), timestep for \textsc{PeTar} ($\Delta t_{\rm soft}$), cut-off radius for \textsc{PeTar} ($r_{\rm out}$).
\end{tabnote}

\end{table*}

\subsubsection{Formation of star cluster complex}
\begin{figure*}
 \begin{center}
  \includegraphics[width=7.8cm]{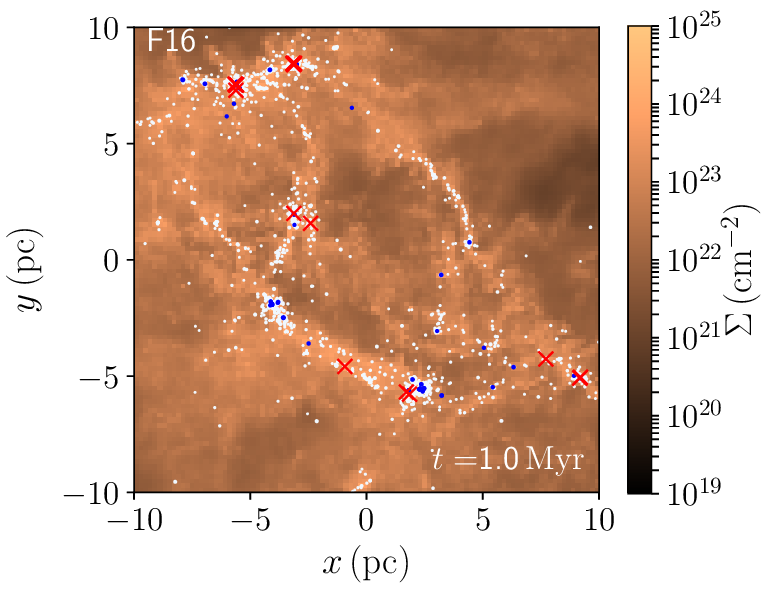}
  \includegraphics[width=6.3cm]{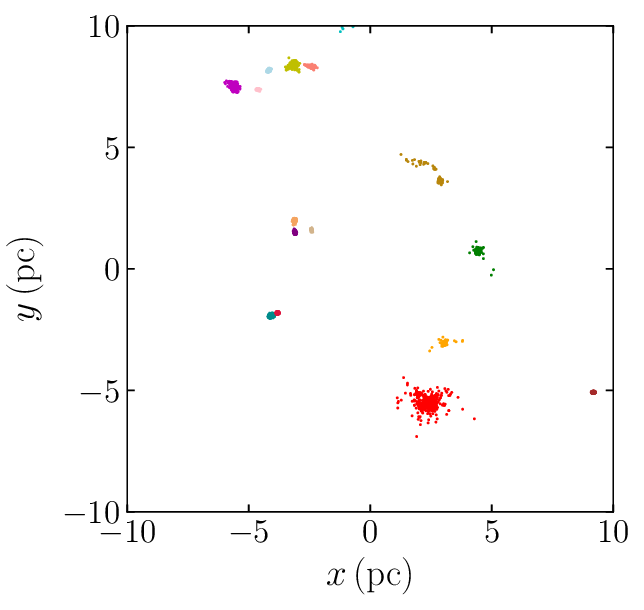}
 \end{center}
\caption{(Left) Snapshot at $t=1$\,Myr for Model F16. White points indicate stars more massive than $1 M_{\odot}$, blue points are massive stars with $>20 M_{\odot}$. Red crosses indicate the positions of binaries with a semi-major axis of $<1000$\,au. (Right) Snapshot of clusters detected using HOP algorithm} at $t=1$\,Myr. \label{fig:F16_snap}
\end{figure*}

\begin{figure}
 \begin{center}
  \includegraphics[width=7.8cm]{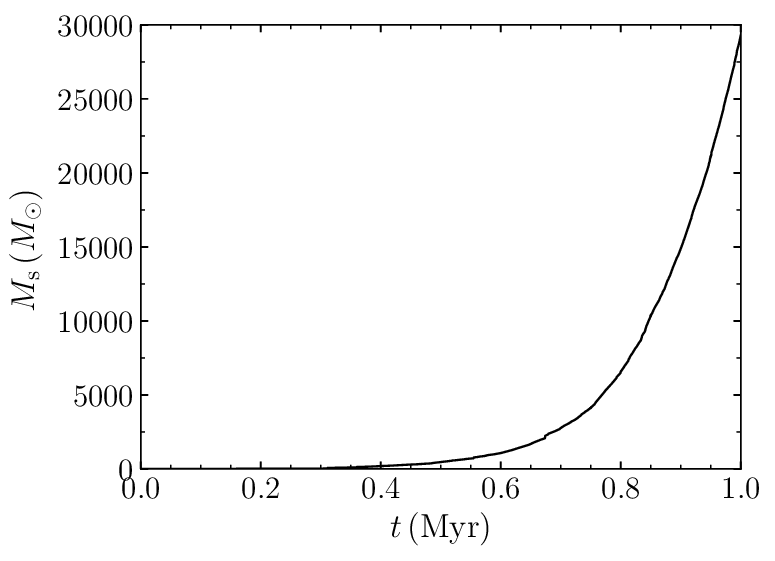}
 \end{center}
\caption{Time evolution of the total stellar mass formed in model F16.}\label{fig:F16_mass}
\end{figure}

In the left panel of Fig.~\ref{fig:F16_snap}, we present the snapshot at 1\,Myr. Similar to the cluster formation model, stars form in dense regions, which are hubs and filaments. 
In Fig.~\ref{fig:F16_mass}, we present the evolution of stellar mass. After about an initial free-fall time ($\sim 0.8$\,Myr), the star formation is accelerated. Since we do not assume any feedback processes from massive stars in this simulation, the star formation does not stop until the gas is fed up by star formation. We therefore stop the simulation at 1\,Myr. 

In our simulation, all massive stars formed inside star clusters (or clumps). This is because the formation of massive star needs sufficient amount of mass in our star formation model. However, some massive stars are located in less dense regions. Such massive stars formed in dense regions and were later ejected due to close encounters with other stars. They are so-called runaway or walkaway stars \citep{1954ApJ...119..689B,2011Sci...334.1380F}. Our code enables the integration of stellar orbits without gravitational softening, and as a consequence, we can simulate the formation of OB runaway stars in simulations of star cluster formation. 
This kind of scattered massive stars would be important when we take feedback from massive stars into account. The ejection of massive stars can weaken the feedback in the central region of star clusters and can delay the gas expulsion time \citep{2020arXiv200309011W, 2020MNRAS.tmp.2473D}.  

\subsubsection{Binary property}
\begin{figure*}
 \begin{center}
  \includegraphics[width=7.8cm]{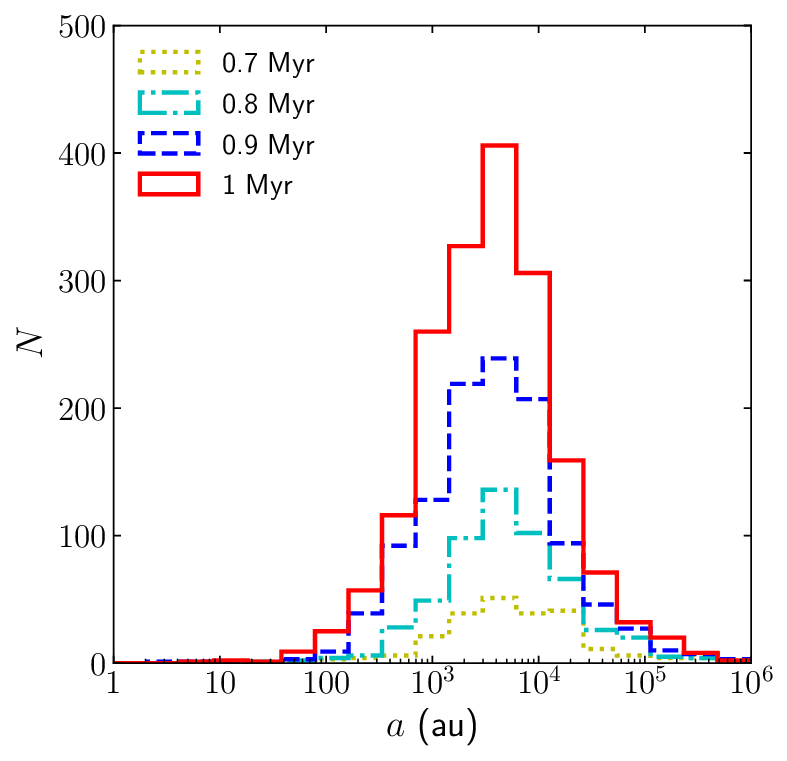}
  \includegraphics[width=7.8cm]{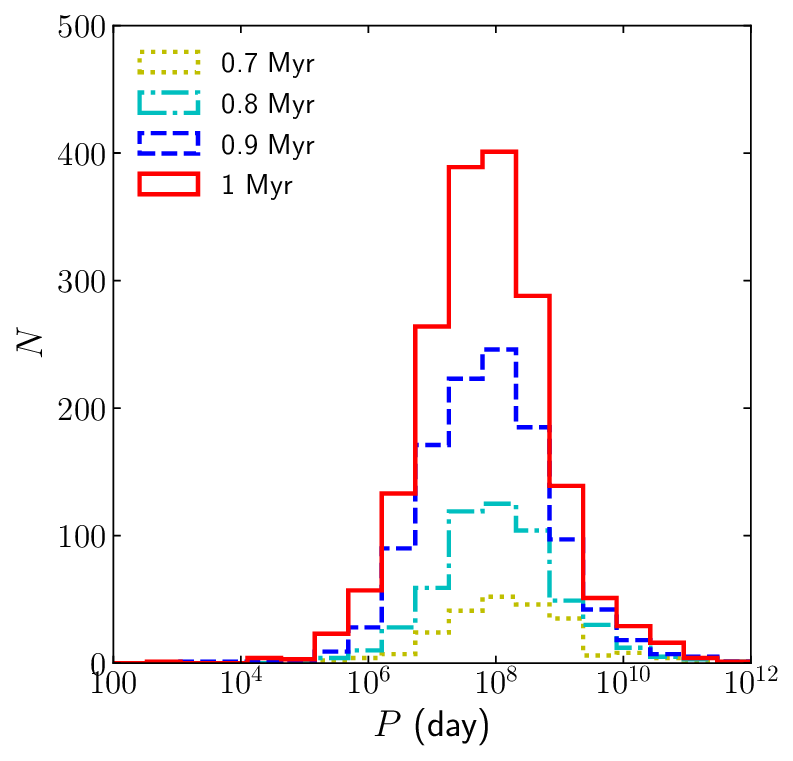}
 \end{center}
\caption{Semi-major axis (left) and period (right) distributions of binaries at 0.7, 0.8, 0.9, and 1\,Myr for model F15.}\label{fig:binary_dist}
\end{figure*}

With \textsc{PeTar}, we can precisely integrate binaries. We detect binaries with a binding energy of $>10^{-3}\,kT$, where $1\,kT$ is the mean kinetic energy of stars in the stellar system, using \textsc{AMUSE}.  Fig.~\ref{fig:binary_dist} presents the distributions of semi-major axes and eccentricities of all detected binaries. 
Only one very hard binary has the semi-major axis smaller than 10\,au and there are several with $<100$\,au. 
We emphasize that this code can integrate the entire system including such tight binaries. However, the majority of our detected binaries are much softer. The distributions of semi-major axes and periods peaks in $10^3$--$10^4$\,au and in $10^7$--$10^9$ days, respectively.

\begin{figure}
 \begin{center}
  \includegraphics[width=7.8cm]{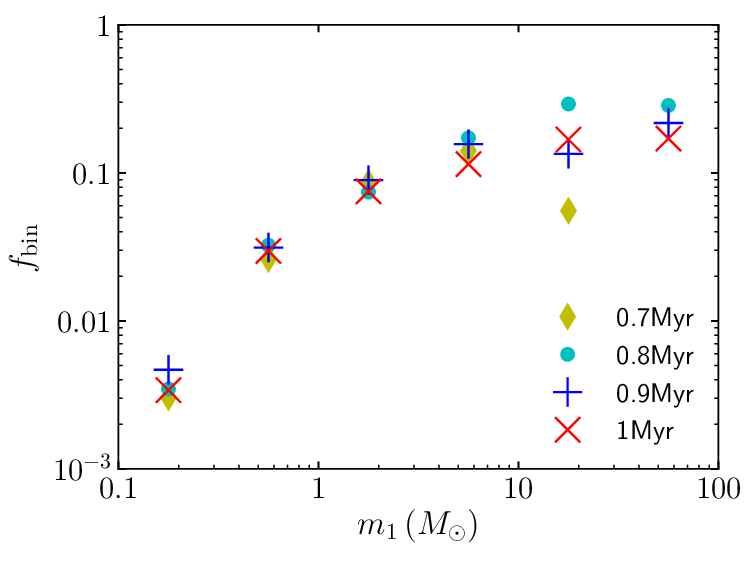}
 \end{center}
\caption{Binary fraction as a function of the primary mass ($m_1$) for model F16 at 0.7, 0.8, 0.9, and 1\,Myr.}\label{fig:binary_frac_m1}
\end{figure}

The number of binaries increases with time as the star formation proceeds. Fig.~\ref{fig:binary_frac_m1} presents the binary fraction as a function of primary mass of binaries ($m_1$). The binary fraction increases from 0.7 to 0.8\,Myr and then decreases in later time. The decrease is probably due to the disruption of soft binaries. 
Our binary fraction shown in the figure is consistent with that of the similar simulations from \citet{2019ApJ...887...62W}, which is $\sim 0.1$ at $\sim 1 M_{\odot}$ and $\sim 0.4$ at $\sim 10 M_{\odot}$). 
\citet{2010MNRAS.404..721M} performed a simulation of star-cluster formation with a smaller and more compact system.
Their semi-major axis distribution of binaries peaks at 1--10\,au, and their multiple fraction ($\sim 0.5$ for stars with $\sim 1 M_{\odot}$) is higher than ours. However, the maximum stellar mass in their simulation is only $10 M_{\odot}$, but $\sim 100 M_{\odot}$ in ours.

The observed binary fraction, $\sim 0.5$ for $\sim 1 M_{\odot}$ and $>0.8$ for $\sim 10 M_{\odot}$ \citep{2013ARA&A..51..269D}, is also higher than ours. However, our simulation reproduces the increase of the binary fraction from low to high mass stars. 
The low binary fraction in our simulation may be due to the lack of the formation of primordial binaries, which should occur in a scale smaller than our resolution limit ($\sim 0.1$\,pc in our simulation). In the future, We will investigate the evolution of binaries in star clusters with primordial binaries.

\begin{figure}
 \begin{center}
  \includegraphics[width=7.8cm]{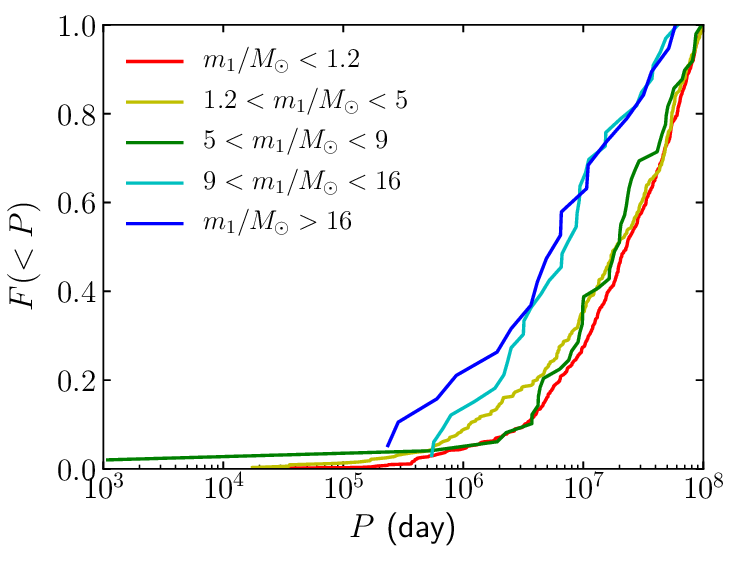}
 \end{center}
\caption{Cumulative distributions of binary orbital periods ($P$) in each primary mass range at 1\,Myr for model F15.}\label{fig:binary_cum_P}
\end{figure}

\begin{figure*}
 \begin{center}
  \includegraphics[width=7.8cm]{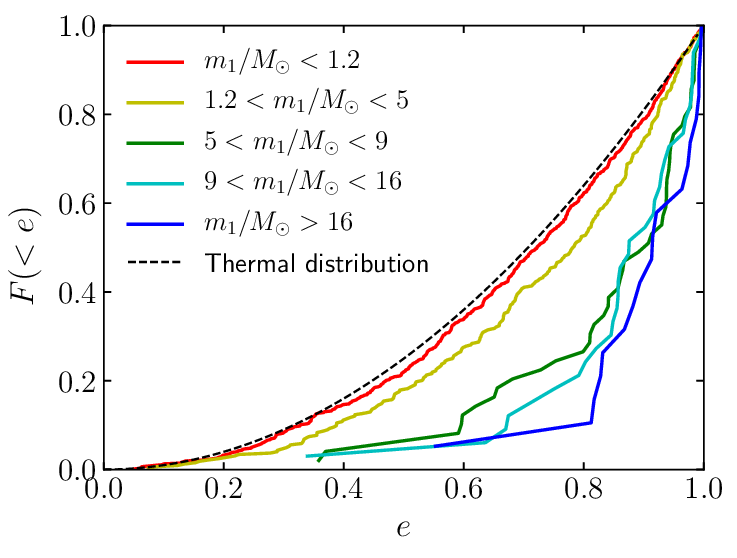}
  \includegraphics[width=7.8cm]{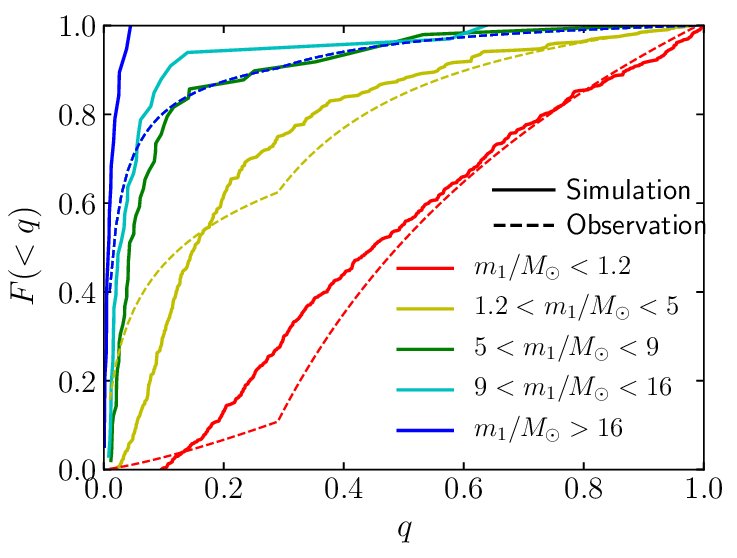}
 \end{center}
\caption{Cumulative distributions of binaries with a period of $10^8$ days in Model F16 at $t=1$\,Myr. (Left) Distribution of eccentricities ($e$). (Right) Distribution of mass ratio ($q$). Filled curves are for the simulation, and dashed curves are for observations ($f(q)\propto q^{\gamma}$, where $\gamma=\gamma_{\rm largeq}$ for $q>0.3$ and $\gamma=\gamma_{\rm smallq}$ for $q<0.3$). From \citet{2017ApJS..230...15M}, $\gamma_{\rm smallq}=0.3$ and $\gamma_{\rm largeq}=-1.1$ for $m_1=0.8$--1.2\,$M_{\odot}$, $\gamma_{\rm smallq}=-1.0$ and $\gamma_{\rm largeq}=-2.0$ for $m_1=2$--5\,$M_{\odot}$, and $\gamma_{\rm smallq}=-1.5$ and $\gamma_{\rm largeq}=-2.0$ for $m_1>5\,M_{\odot}$ for long-period binaries with $\sim 10^7$ days.}\label{fig:binary_cum_dist}
\end{figure*}

Fig.~\ref{fig:binary_cum_P} presents the cumulative distribution of the orbital periods of binaries in our simulation for each primary-mass range. 
We set the maximum period to be $10^8$ days, which is observational limit \citep{2017ApJS..230...15M}. 
The result indicates that the binary period is shorter for more massive stars. 
Fig.~\ref{fig:binary_cum_dist} presents the cumulative distributions of binaries as a function of binary eccentricity ($e$) and mass ratio ($q$) in each primary-mass range. The eccentricity distribution of low-mass stars is thermal. As the primary mass increases, more eccentric binaries appear. The mass-ratio distribution also depends on the primary masses. In the figure, we also plot the mass-ratio distribution of long-period binaries obtained from the observations \citep{2017ApJS..230...15M}. In both the observations and the simulations, massive stars tend to have small $q$.

The spatial distribution of relatively tight binaries with a semi-major axis of $10^3$\,au is shown in Fig.~\ref{fig:F16_snap}. They are inside clumps. Since we did not assume any primordial binaries, they all dynamically formed inside the clumps. From the relation between primary mass ($m_1$) and mass ratio ($q$), we find more massive stars have a smaller mass ratio. This is a statistical result because each clump is still small and therefore it does not contain multiple massive stars. Observationally, on the other hand, massive stars have a high binary fraction and the companion mass is close to the primary mass \citep{2012Sci...337..444S}. Some clumps include multiple massive stars. In such clumps, massive hard binaries may be able to form due to the mass segregation and repeating three-body encounters. However, the integration time of $<1$\,Myr is too short to discuss the mass segregation. In addition, we find no binaries with a period of $\lesssim 1000$ days. Such shorter-period binaries may be primordial.

\subsubsection{Formed clusters}

We detect star clusters using the HOP algorithm \citep{1998ApJ...498..137E} on \textsc{AMUSE}, which is an algorithm to separate sub-structures based on their local stellar densities. We use a parameter set that is the same as those in \citet{2019MNRAS.486.3019F}. We detect 25 clumps which includes more than 100 stars. In the right panel of Fig.~\ref{fig:F16_snap}, we present the spatial distribution of the detected clumps but only for those with more than 200 stars. 

\begin{figure}
 \begin{center}
  \includegraphics[width=7.8cm]{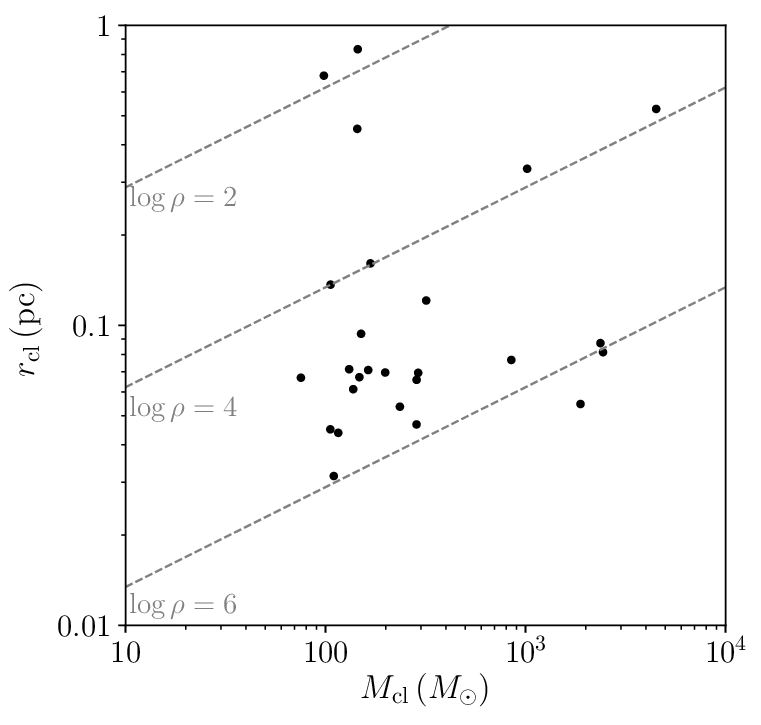}
 \end{center}
\caption{Relation between masses and half-mass radii of detected clusters for model F16 at $t=1$\,Myr. Gray dashed lines indicate the densities of $10^2$, $10^4$, and $10^6 M_{\odot}$\,pc$^{-3}$ from top to bottom.}\label{fig:cluster_mass_rad}
\end{figure}

We also present the mass-radius relation of these clusters in Fig.~\ref{fig:cluster_mass_rad}. The detected clumps have a mass range similar to observed open clusters, but the density of the detected clusters is one or two-order of magnitude higher than the observed ones \citep[see figure~2 in ][]{2010ARA&A..48..431P}. Including any feedback, the density later will drop due to the gas expulsion. This implies that star clusters were born with a density much higher than currently observed ones and became less dense after the gas expulsion. We further investigate the evolution of star clusters over gas expulsion using a code with feedback from massive stars and discuss the effect of the feedback in \citet{2021arXiv210302829F}.

\section{Summary}

We developed \textsc{ASURA+BRIDGE}, a tree-direct hybrid $N$-body/SPH code as a part of the SIRIUS project \citep[see also][]{2020arXiv200512906H}. In \textsc{ASURA+BRIDGE}, stellar particles are integrated using a sixth-order Hermite code or the \textsc{PeTar} code \citep{Wang2020c}. The latter combines P$^3$T \citep{2011PASJ...63..881O, 2015ComAC...2....6I} and SDAR methods \citep{Wang2020b} so that close encounters and orbits of binaries can be accurately evolved. Therefore, \textsc{ASURA+BRIDGE} can treat stellar dynamics properly without any gravitational softening. On the other hand, gas particles are integrated using \textsc{ASURA}, which is an SPH code including cooling/heating models \citep{2008PASJ...60..667S,2009PASJ...61..481S}.

We performed some test simulations of star cluster models embedded in gas clouds. We found that the timestep for Bridge must be shorter than the local free-fall time of gas, i.e., the timestep for gas particles. If the Bridge timestep is larger than the critical value, the stellar distribution drifts. 

Thanks to the zero-softening treatment, we can properly follow the dynamical evolution of star clusters. With softening, stars form an artificial dense small core in the cluster center. Without softening, on the other hand, stars are scattered due to close encounters and as a consequence, the artificial core disappears. 

We also tested a model with star formation using a statistical manner \citep{2020arXiv200512906H}. We performed a series of simulations for initially homogeneous molecular clouds with turbulent velocity fields. Although the total mass of the forming clusters depends on the random seed for the initial turbulence, our results are consistent with a similar simulation using a sink-particle scheme. 

The dependence of our results on the resolution is also tested. The total mass of the forming stars slightly depends on the resolution, i.e., simulations with higher resolution result in the formation of more stars. The total stellar mass does not strongly depend on the local star formation rate which we assume for our star formation scheme. This is because the threshold density for star formation is sufficiently high and the local dynamical time (free-fall time) is short enough compared with the dynamical timescale of the entire system \citep{2008PASJ...60..667S}.

The radial density distribution of the formed star clusters does not strongly depend on the mass resolution of gas either, but slightly depends on the softening length of gas. This result suggests that we can adopt a relatively low resolution of gas ($m_{\rm gas}\sim 0.1\,M_{\odot}$) for this kind of simulations if we are interested in the structure and dynamical evolution of star clusters formed in relatively large-scale turbulent clouds. Our scheme would also be applicable to larger-scale simulations such as globular clusters and galaxies by adopting a resolution of $0.1$--$1 M_{\odot}$. 

We finally performed a simulation for the formation of a star cluster complex, which includes multiple clusters. In this simulation, the final stellar mass reached $3\times 10^4M_{\odot}$, which we aimed to achieve. With \textsc{PeTar}, we can integrate stars with $O(N\log N)$ and treat binary using an SDAR method \citep{Wang2020b}. We obtained a fraction of binaries in our simulation and we confirmed that our binary fraction is consistent with those in previous similar studies \citep{2010MNRAS.404..721M,2019ApJ...887...62W}, but lower than observation \citep{2013ARA&A..51..269D}. This may be due to the lack of primordial binaries in our current scheme. The binaries detected in our simulation mainly distribute at the orbital period of $10^7$--$10^9$ days, which corresponds to long-period binaries in observations. Our binary fraction increased as the primary mass increased. This is also consistent with previous studies in both simulations and observations.

We stopped our simulation at 1\,Myr ($1.2$ initial free-fall time) since we do not include feedback from massive stars and therefore star formation cannot be suppressed. 
We will include feedback in the code, continue the simulations for a longer time, and discuss these points in our next paper \citep{2021arXiv210302829F}.

\begin{ack}
The authors thank the anonymous referee for the useful comments.
Numerical computations were carried out on Cray
XC50 CPU-cluster at the Center for Computational Astrophysics (CfCA) of the National Astronomical Observatory of Japan and Oakbridge-CX at Information Technology Center, The University of Tokyo. 
This work was supported by JSPS KAKENHI Grant Number 19H01933, 20K14532 and Initiative on Promotion of Supercomputing for Young or Women Researchers, Information Technology Center, The University of Tokyo, and MEXT as “Program for Promoting Researches on the Supercomputer Fugaku” (Toward a unified view of the universe: from large scale structures to planets, Revealing the formation history of the universe with large-scale simulations and astronomical big data).
MF was supported by The University of Tokyo Excellent Young Researcher Program. YH was supported by the Special Postdoctoral Researchers (SPDR) program at RIKEN.

\end{ack}

\bibliography{reference}

\end{document}